\documentclass[journal=ancac3,manuscript=article]{achemso}

\usepackage[T1]{fontenc} 
\usepackage[utf8x]{inputenc}
\usepackage{xcolor}
\usepackage{amsmath}

\def\eps{\epsilon}

\author{Arturo Moncho-Jordá}
\affiliation{Instituto Carlos I de Física Teórica y Computacional, Facultad de Ciencias, Universidad de Granada, Avenida Fuentenueva S/N, 18001 Granada, Spain}
\alsoaffiliation[Universidad de Granada]{Departamento de Física Aplicada, Facultad de Ciencias, Universidad de Granada, Avenida Fuentenueva S/N, 18001 Granada, Spain}
\email{moncho@ugr.es}
\author{Alicia Germán-Bellod}
\affiliation[Universidad de Granada]{Departamento de Física Aplicada, Facultad de Ciencias, Universidad de Granada, Avenida Fuentenueva S/N, 18001 Granada, Spain}
\author{Stefano Angioletti-Uberti}
\affiliation[Imperial College London]{Department of Materials, Imperial College London, London, UK}
\author{Irene Adroher-Benítez}
\affiliation[SISSA]{Scuola Internazionale Superiore di Studi Avanzati (SISSA), Trieste, Italy}
\author{Joachim Dzubiella}
\affiliation[Helmholtz-Zentrum Berlin f\"ur Materialien und Energie]{Research Group for Simulations of Energy Materials, Helmholtz-Zentrum Berlin f\"ur Materialien und Energie, Hahn-Meitner-Platz 1, D-14109 Berlin, Germany}
\alsoaffiliation[Albert-Ludwigs-Universit\"at Freiburg]{Physikalisches Institut, Albert-Ludwigs-Universit\"at Freiburg, Hermann-Herder Str. 3, D-79104 Freiburg, Germany}
\email{joachim.dzubiella@physik.uni-freiburg.de}

\title{Non-Equilibrium Uptake Kinetics of Molecular Cargo into Hollow Hydrogels Tuned by Electrosteric Interactions}


\begin{document}

\begin{tocentry}

%
%
%

\centering
\includegraphics[height=3.5cm]{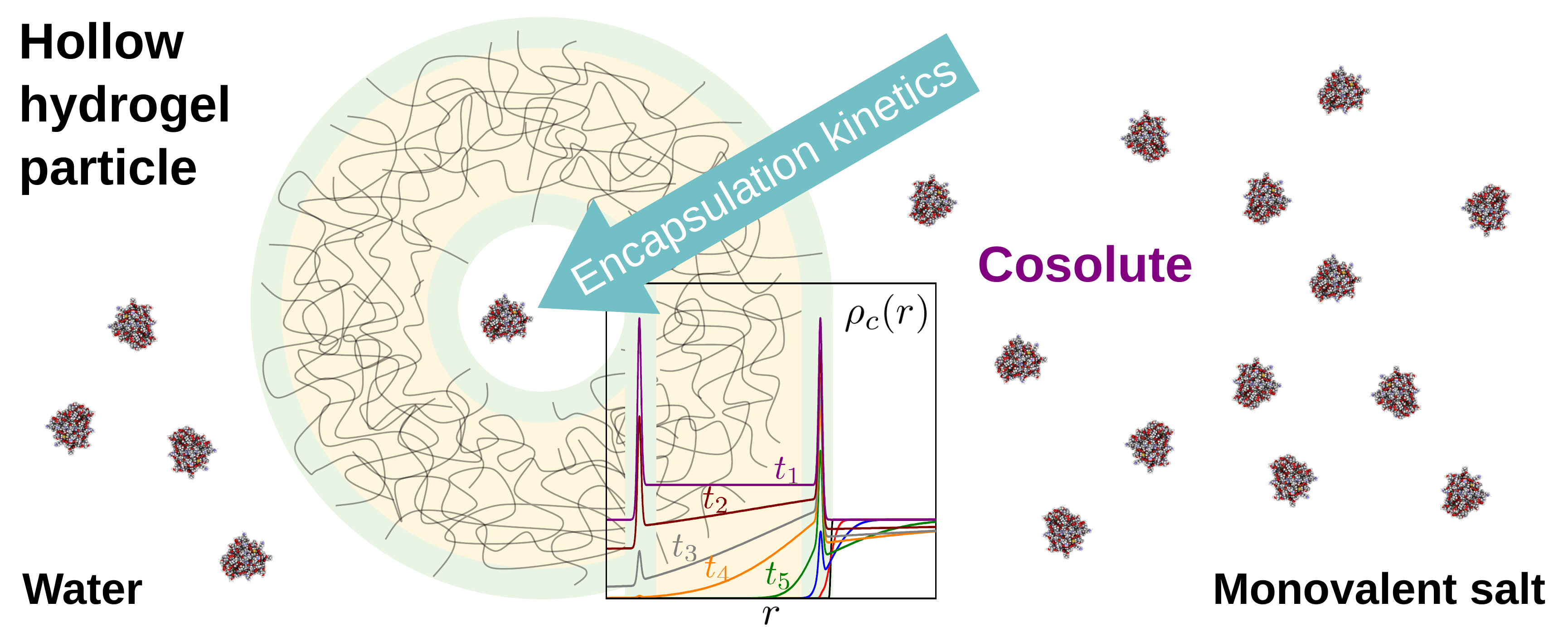}

\end{tocentry}

\begin{abstract}
Hollow hydrogels represent excellent nano- and micro-carriers due to their ability to encapsulate and release large amounts of cargo molecules (cosolutes) such as reactants, drugs, and proteins. In this work, we use a combination of a phenomenological effective cosolute-hydrogel interaction potential and Dynamic Density Functional Theory to investigate the full non-equilibrium encapsulation kinetics of charged and dipolar cosolutes by an isolated charged hollow hydrogel immersed in a 1:1 electrolyte aqueous solution. Our analysis covers a broad spectrum of cosolute valences ($z_\textmd{c}$) and electric dipole moments ($\mu_\textmd{c}$), as well as hydrogel swelling states and hydrogel charge densities. Our calculations show that, close to the collapsed state, the polar cosolutes are predominantly precluded and the encapsulation process is strongly hindered by the excluded-volume interaction exerted by the polymer network. Different equilibrium and kinetic sorption regimes (interface {\it versus} interior) are found depending on the value and sign of  $z_\textmd{c}$ and the value of $\mu_\textmd{c}$. For cosolutes of the same sign of charge as the gel, the superposition of steric and electrostatic repulsion leads to an "interaction-controlled" encapsulation process, in which the characteristic time to fill the empty core of the hydrogel grows exponentially with $z_\textmd{c}$. On the other hand, for cosolutes oppositely charged to the gel, we find a "diffusion-controlled" kinetic regime, where cosolutes tend to rapidly absorb into the hydrogel membrane and the encapsulation rate only depends on the cosolute diffusion time across the membrane. Finally, we find that increasing $\mu_\textmd{c}$ promotes the appearance of metastable and stable surface adsorption states. For large enough $\mu_\textmd{c}$, the kinetics enters a "adsorption-hindered diffusion", where the enhanced surface adsorption imposes a barrier and slows down the uptake. Our study represents the first attempt to systematically describe how the swelling state of the hydrogel and other leading physical interaction parameters determine the encapsulation kinetics and the final equilibrium distribution of polar molecular cargo.
\end{abstract}



Hydrogels are soft nanoparticles formed by a cross-linked polymer network immersed in water. During the last decades, hydrogels have gained considerable attention from both, theoretical and experimental points of view, due to their ability of swelling in response to many external stimuli~\cite{Murray1995,Saunders2009,Stuart2010}. A typical example of a thermoresponsive hydrogel is the one formed by poly(N-isopropylacrylamide) (pNIPAM), whose phase behavior is nowadays well understood~\cite{Gilcreest, Halperin2018,Gorelov1997}. Swelling is a reversible process, allowing the reuse of the particles over many swelling/shrinking cycles. The presence of cross-links provides structural integrity, so hydrogel particles do not dissolve in the swelling process. In addition, hydrogels have a high permeability, and can be designed to be biocompatible and biodegradable. These important features make hydrogels very promising biomaterials for many versatile applications~\cite{Stefano2017}. For instance, hydrogels represent excellent nanocarriers to incorporate and release host biomolecules or drugs in a responsive fashion. In the particular case of drug delivery, they can be used to control the dose, release rate and location of the therapeutic drug, acting as a protecting cover of the encapsulated molecule when immersed in biological environments~\cite{Kim1992,Ghugare2009,Vinogradov2010,Bae2013}. They can also be used in biosensing applications~\cite{Goff2015}, as nano or microreactors with tuneable catalytic reaction rates~\cite{Wu2012,Jia2016} or as selective traps for chemical separation and purification~\cite{Menne2014,Kureha2018}.

Of particular interest to some applications are stimuli-responsive hollow hydrogels: particles formed by a single or multiple cross-linked polymeric shells with a hole inside, in which the swelling of the polymeric network may be varied, for instance, with temperature of solvent pH~\cite{Nayak2005,Schmid2016,Contreras-Caceres2015,Brugnoni2018}. On the one hand, the presence of the internal void greatly enhances the load capacity of different kind of molecules (drugs, proteins and other biomolecules, reactants,...)~\cite{Bedard2010,Yan2010,Chiang2013,McMasters2017}. On the other hand, by tuning the swelling state of the hydrogel membrane, the diffusion process and, consequently, the cargo encapsulation/release rate, can be externally manipulated~\cite{Masoud2012}. In fact, if the hollow hydrogel approaches the impermeable collapsed state, cargo molecules trapped in the void become completely isolated from the bulk suspension. Inducing progressive swelling states, the cargo release rate can be gradually increased~\cite{Motornov2011,Xing2011}. For instance, hollow hydrogels of poly(4-vinylpyridine) have been recently used in {\it in-vivo} experiments to encapsulate paclitaxel molecules (one of the most effective cytotoxins for the treatment of breast and lung cancer) to increase antitumoral efficacy accompanied by efficient cell internalization and reduction of collateral effects~\cite{Contreras-Caceres2017}. In addition, poly(thyol-ene) and poly(methacrylic anhydride) microcapsules have been successfully employed to encapsulate neutral and charged molecules of different molecular weights, adjusting the cargo release on demand. The permeability of these microcapsules was strongly dependent on the degree of swelling, and could be actively and dynamically modified by varying the pH of the medium~\cite{Amato2017,Werner2018,Werner2018b}.

Controlling the kinetics of the encapsulation/release process is one of the major goals in all these applications. In some situations, a high release rate is required to achieve a fast response after the trigger stimulus. In other applications, a slow and gradual cargo release of drugs, proteins and other bioactive agents is more convenient to achieve sustained effects. Therefore, understanding the physicochemical interactions involved in the kinetics of this kind of processes is fundamental to improve the efficiency of newly-designed hydrogels.

However, in spite of the large number of potential applications, a complete theoretical description of the equilibrium and dynamic (non-equilibrium) transport properties of hollow hydrogels employed in cargo encapsulation or release is still lacking. In this regard, it is remarkable that most models to describe the encapsulation or release kinetics of these systems are still based on the ideal diffusion equation~\cite{Higuchi1963,Ritger1987,Ritger1987a,Korsmeyer1981,Lee2011,Li2011}. More precisely, the coupling introduced by the interactions between the matrix and the drug is not explicitly considered and their effects are lumped together in the value of effective constants such as the drug diffusivity or solubility. These approximations were already questioned by other authors. Petropoulos, Papadokostakis and Amarantos~\cite{Petropoulos1992,Petropoulos2012}, for instance,  showed that the kinetics of drug release can be written as a generalised diffusion equation, where the activity of the various components appear. The latter can be extrapolated from fitting equilibrium adsorption isotherms and used to calculate the kinetics more reliably~\cite{Papadokostaki2002,Papadokostaki2004,Papadokostaki2004a}. A more fundamental and predictive approach can be built instead by recognising that the activity simply measure non-ideal behaviour, hence its effects can be directly calculated from the microscopic interactions in the system. This is exactly the route we take here, and in doing this we show how these underlying physical interactions between the molecule and the polymer network play a key role in controlling the encapsulation rate and the location of the uploaded cargo. For example, our theory predicts that depending on these interactions the cargo can mainly partition in three different ways: inside the internal core, into the hydrogel membrane, or superficially adsorbed onto the external layers of the hollow hydrogel. Not surprisingly, the encapsulation kinetics also depends on the diffusion coefficient of the molecule inside the polymer network. Indeed, obstruction and hydrodynamic retardation effects exerted by the polymers, considered in our framework, can drastically slow down the diffusion across the hydrogel membrane, especially for shrunken hydrogels~\cite{Amsden1998,masaro1999physical,Matej2018}.


The aim of this paper is to provide a theoretical background to predict the full non-equilibrium encapsulation kinetics and the final equilibrium distribution of a certain neutral or charged substance (which we will refer as cargo or cosolute) through the hydrogel membrane for many distinct conditions (swelling state and density charge of the hydrogel, and charge and electric dipolar momentum of the cargo). For this purpose, a phenomenological effective pair potential, $V_\textmd{eff}(r)$, that combines electrostatic, osmotic and excluded-volume (steric) energetic contributions is deduced for the cosolute-hydrogel, interactions~\cite{Adroher-Benitez2017}. In our model, we consider the general situation of non-uniformly charged cosolutes (such as the case of heterogeneous reactants, ligands, small proteins, {\it etc}), and include the Born solvation attraction to account for the charge screening effects caused by the excess of counterions inside the charged hydrogel. We employ the resulting effective Hamiltonian to explore a wide spectrum of parameters, and summarize the results by means of state diagrams, where the final equilibrium states are classified into seven different categories: weak and strong exclusion/absorption of cosolute inside the hydrogel membrane, and metastable, weak and strong adsorption onto the surfaces of the hydrogel.  

Using this effective hydrogel-cosolute interaction, we propose an equilibrium density functional, and then generalize it to non-equilibrium situations by means of the theoretical formalism called Dynamic Density Functional Theory (DDFT)~\cite{Marconi1999}, which has been successfully applied to similar problems~\cite{Angioletti-Uberti2014,Angioletti-Uberti2018}. The resulting theory is then employed to obtain the time evolution of the cosolute concentration in terms of the hydrogel-cosolute interactions and its local diffusion coefficient inside the hydrogel network for different cosolute charge, dipole moments and swelling states. As the presented model includes the excluded-volume effects of the effective interaction and cosolute diffusivity, it can be applied to any swelling state of the hydrogel, from the collapsed to the swollen conformations.

\section{Results and discussion}

\begin{figure}[ht!]
	\centering
	\includegraphics[width=0.5\linewidth]{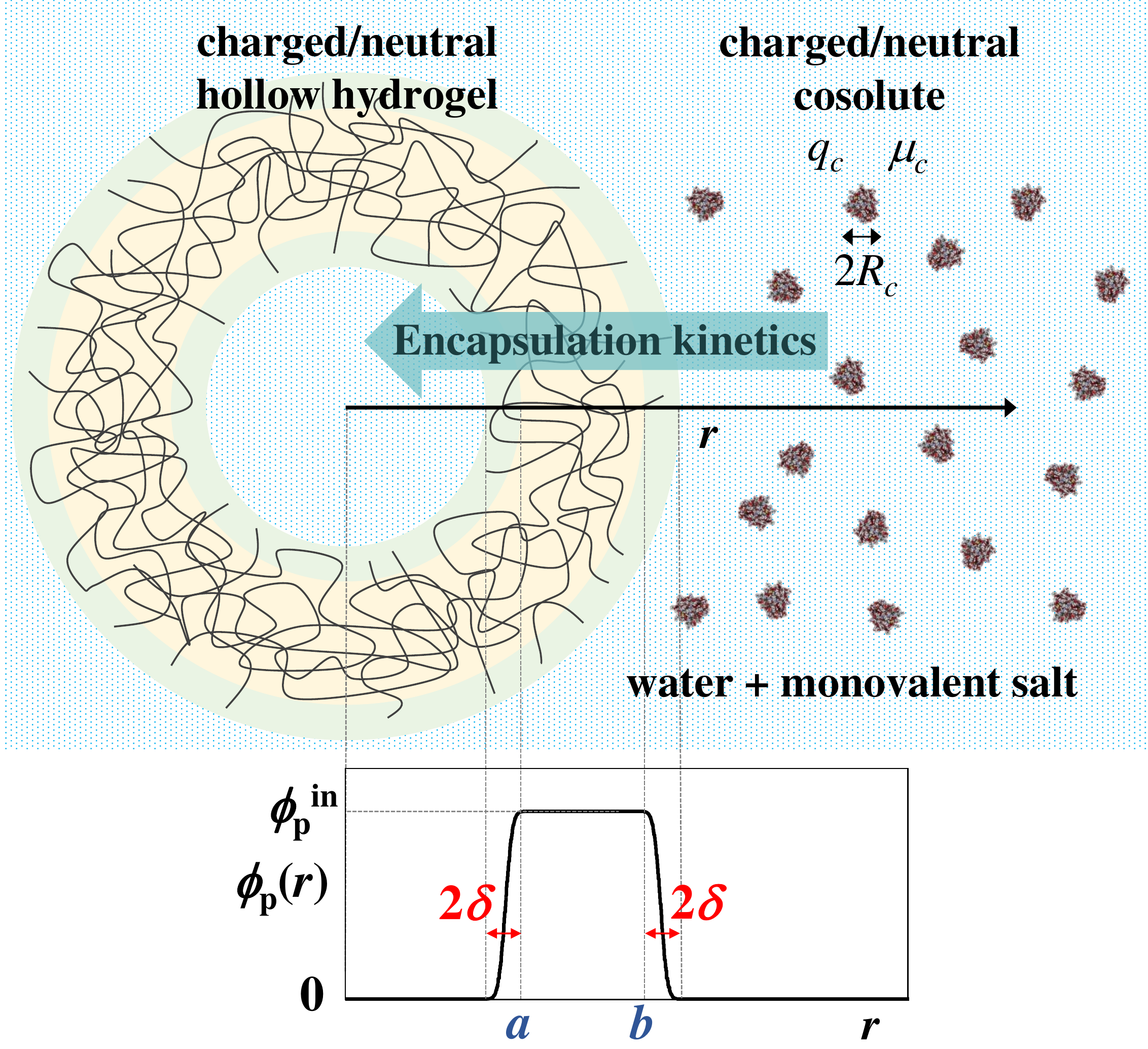}
	\caption{Schematic illustration of the system: a single charged hollow hydrogel immersed in a salty solution with charged cosolute particles of radius $R_\textmd{c}$. Cosolutes with a typical diameter smaller than the mesh size diffuse from the external volume to the internal one, leading to an encapsulation process. Below, the polymer volume fraction is plotted as a function of the distance to the hydrogel center, $r$.}
	\label{fig:fig1}
\end{figure}

The system under study consists of a single hollow hydrogel particle immersed in an aqueous solution of monovalent salt at concentration $\rho_\textmd{s}$ with a certain amount of cosolute molecules suspended in the bulk solution, at concentration $\rho_\textmd{c}^\textmd{bulk}$. Ions are assumed to be point-like, and the solvent (water) is treated in our model as a uniform background of relative dielectric constant $\epsilon_r=78.5$. We consider that the hydrogel is spherical, with internal and external radius given by $a$ and $b$, respectively. The polymer density inside real hydrogels does not drop abruptly to zero at the interface. This is especially true for swollen hydrogels, where the polymer density is more or less uniform inside the network, but gradually decreases to zero at the external interface~\cite{Fernandez-Barbero2002}. To model this radial dependence for the hollow hydrogel, we combine two error functions. Therefore, the polymer volume fraction is given by
\begin{equation}
\label{phipr}
\phi_p(r)=\frac{\phi_\textmd{p}^\textmd{in}}{2}\left[ \textmd{erf} (2(r-R_1)/\delta) -\textmd{erf} (2(r-R_2)/\delta) \right],
\end{equation}
where $r$ is the distance to the hydrogel center, $\phi_\textmd{p}^\textmd{in}$ is the polymer volume fraction inside the spherical membrane, $2\delta$ represents the thickness of both, the internal and external interfaces, $R_1=a-\delta$ and $R_2=b+\delta$ (see illustration in Figure~\ref{fig:fig1}). The concentration of charged monomers inside the hydrogel is supposed to follow the same profile than the polymer volume fraction
\begin{equation}
\rho_\textmd{m}(r)=\frac{\rho_\textmd{m}^\textmd{in}}{\phi_\textmd{p}^\textmd{in}}\phi_\textmd{p}(r),
\end{equation}
where $\rho_\textmd{m}^\textmd{in}$ is the concentration of charged monomers inside the hydrogel membrane.

The swelling ratio of the hydrogel, $q$, is a crucial parameter, as it controls the equilibrium distribution of cosolute molecules and their encapsulation/release kinetics. It is defined as the ratio between the radius at the swollen state, $b$, and at the reference collapsed state, $b_0$, $q=b/b_0$. We assume uniform swelling, so $a=qa_0$, $\delta = q\delta_0$, $\phi_\textmd{p}^\textmd{in} =q^{-3}\phi_\textmd{p}^{\textmd{in},0}$ and $\rho_\textmd{m}^\textmd{in} =q^{-3}\rho_\textmd{m}^{\textmd{in},0}$ where $a_0$, $\delta_0$, $\phi_\textmd{p}^{\textmd{in},0}$ and $\rho_\textmd{m}^{\textmd{in},0}$ are the corresponding values in the collapsed state.

Regarding the cosolute particles, we do not specify any particular molecule. In fact, our generic cosolute may represent different types of neutral or charged molecules, such as proteins, biomolecules, drugs or chemical reactants. In principle, we only consider the cosolute as a particle of characteristic radius $R_\textmd{c}$, net charge $q_\textmd{c}=z_\textmd{c}e$ (where $e$ is the elementary charge and $z_\textmd{c}$ the cosolute valence) and electric dipole moment $\mu_\textmd{c}$. This dipole moment is necessary in order to account for the electrostatic effects that arise when the cosolute is not uniformly charged. The cosolute concentration at any time $t>0$ at distance $r$ from the hydrogel center is denoted by $\rho_\textmd{c}(r,t)$. Upon reaching the equilibrium state, the final density profile, $\rho_\textmd{c}^\textmd{eq}(r)$, will depend on the external field created by the hydrogel and the ionic cloud.

In this work we make use of a simple model for the effective hydrogel-cosolute interaction potential that gathers an electrostatic term, an osmotic repulsion exerted by the excess of counterions inside the charged hydrogel network and a steric (excluded-volume) repulsion induced by the polymer chains:
\begin{equation}
\label{Veff}
V_\textmd{eff}(r)=V_\textmd{elec}(r)+V_\textmd{osm}(r)+V_\textmd{steric}(r).
\end{equation}
In addition, the electrostatic term is split into three more contributions: {\it (i)} The monopolar interaction due to the coupling between the cosolute charge ($z_\textmd{c}$) and the electrostatic potential generated by the charged hydrogel. {\it (ii)} The effective dipolar attraction, $V_\textmd{dip}$, caused by the coupling between the cosolute electric dipole moment ($\mu_\textmd{c}$) and the electric field induced by the hydrogel (only for polar cosolutes). {\it (iii)} The Born solvation attraction, $\Delta V_\textmd{Born}$, that accounts for the self-energy difference of charging the cosolute inside the hydrogel network with respect to the bulk suspension. Each contribution plays a well-defined role on the cosolute partitioning. For instance, the osmotic and steric terms cause the cosolute exclusion from the polymer network, the Born attraction promotes the its internal absorption, the monopolar term induces internal absorption for oppositely charged cosolutes and exclusion for likely charged, whereas the dipolar attraction emphasizes the adsorption of the polar cosolute onto the external and internal interfaces of the hollow hydrogel. More details about all these energetic contributions to $V_\textmd{eff}$ are given in the Methods Section.


\subsection{Numerical implementation and choice of parameters}

Given the large number of variables involved in the system, we fixed some of them to reduce the parameter space. The Bjerrum length involved in the electrostatic interactions is the one for water at room temperature, $\lambda_B=0.71$~nm. The bulk concentration of monovalent salt is in all cases $\rho_\textmd{s}=0.1$~M. The bulk cosolute concentration used as initial state of the encapsulation process is taken to be $\rho_\textmd{c}^\textmd{bulk}=0.001$~M. The cosolute radius is fixed to $R_\textmd{c}=1$~nm.

The internal and external radius of the hollow hydrogel in the collapsed stated are given by $a_0=30$~nm and $b_0=50$~nm, respectively. The thickness of the interface in the collapsed state is usually very narrow, of about few nanometers~\cite{Berndt2005}. In this work we assume that this thickness is given by $2\delta_0=2$~nm. It is well-known that the hydrogel network is not completely dry in the collapsed state, but instead still holds certain amount of water depending on the nature of the constituent polymers and the morphology of the network. Here, a polymer volume fraction in the collapsed state of $\phi_p^{\textmd{in},0}=0.5$ is used.

We assume that all monomers inside the hydrogel matrix have the same radius, given by $R_\textmd{m}=0.35$~nm. Charged monomers are considered negatively charged, with valence $z_\textmd{m}=-1$. Two possible hydrogel charge densities (in the collapsed state) are investigated, $\rho_\textmd{m}^{\textmd{in},0}=0.1$~M (weakly charged), and $0.5$~M (moderately charged). For each one of these choices, the cosolute charge is varied from being likely charged to oppositely charged compared to the gel, {\it i.e.} in the range $-20 \leq z_{c} \leq +20$, and the cosolute dipole electric moment is changed in the interval $0 \leq \mu_{c} \leq 1000$~D. The broadness of these intervals are large enough to cover almost all the possible cosolutes that
may be involved in the applications, including highly charged and polar proteins~\cite{Berman2000,Felder2007}. The effect of the steric exclusion is explored by varying the swelling ratio from $q=1$ (collapsed) to $q=3$ (swollen). All the dimensions of the hollow hydrogel, polymer volume fraction and charge density are accordingly rescaled for any swollen state.

\subsection{Effective interaction and state diagrams}

\begin{figure}[ht!]
  \centering
  \includegraphics[width=0.5\linewidth]{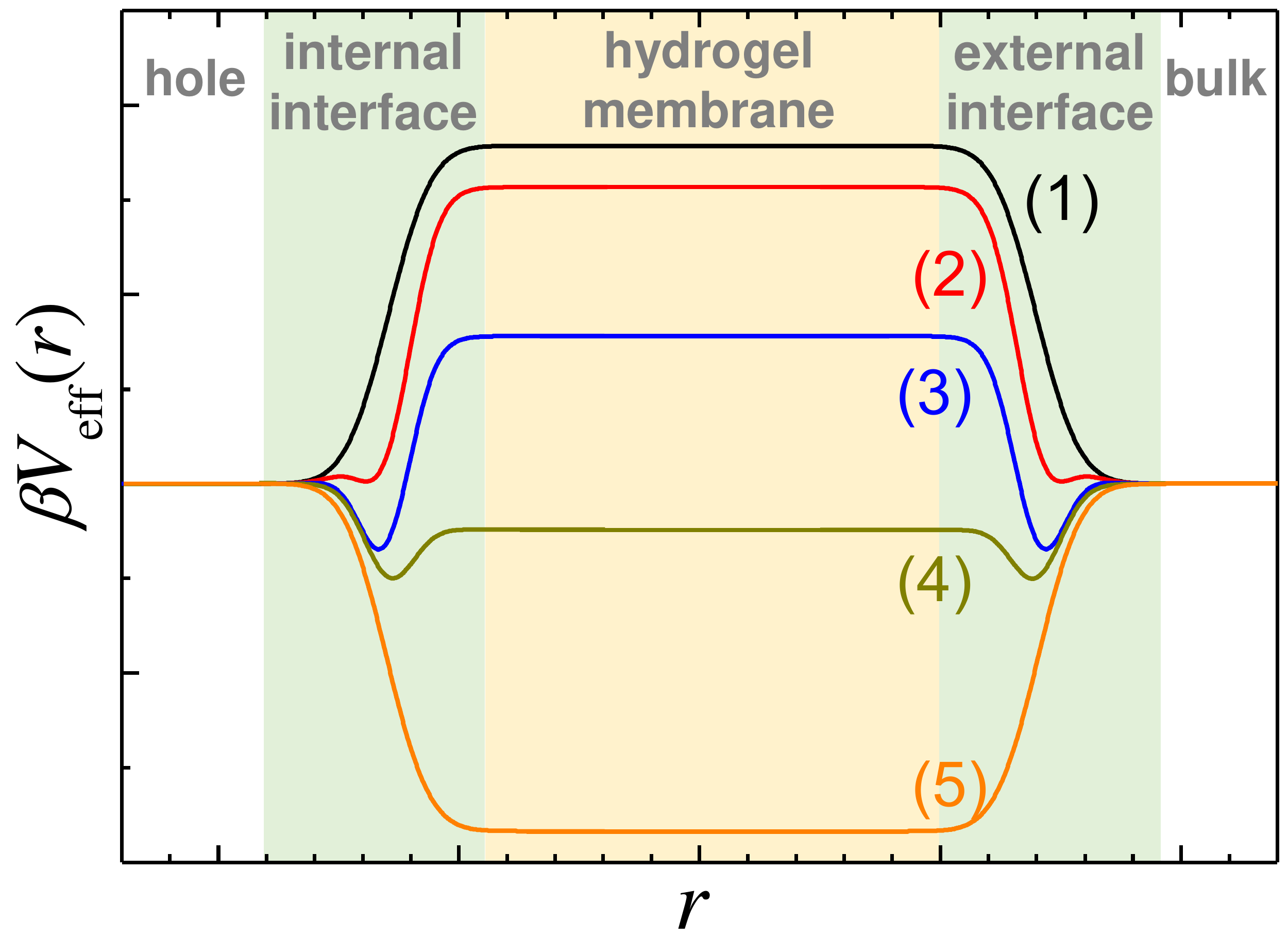}
  \caption{Five different possible situations for the hydrogel-cosolute effective pair potential. (1) Exclusion inside hydrogel network. (2) Exclusion with metastable surface adsorption. (3) Exclusion with stable surface adsorption. (4) Partitioning between absorption inside the hydrogel network and surface adsorption. (5) Network internal absorption.}
  \label{fig:fig2}
\end{figure}

From the shape of the effective potential, it is possible to determine the degree of sorption and predict where the cosolute will preferentially partition. Figure~\ref{fig:fig2} illustrates five different possibilities for $V_\textmd{eff}(r)$, that range from exclusion to absorption inside the hydrogel network, passing through stable or metastable surface adsorption states. In addition, depending on the value of the potential barrier/well, the absorption/adsorption can be strong or weak. In order to get a clear and practical identification of these different regimes, we consider a two-fold classification in terms of the value of $V_\textmd{eff}$ in two regions: inside the hydrogel membrane and at the interfaces. Exclusion is considered \textit{strong} whenever the effective potential inside the hydrogel membrane is $V_\textmd{eff}^\textmd{in}>1$~$k_BT$, and \textit{weak} when $0<V_\textmd{eff}^\textmd{in}<1$~$k_BT$. Analogously, the internal absorption is assumed to be strong for $V_\textmd{eff}^\textmd{in}<-1$~$k_BT$, and weak when $0>V_\textmd{eff}^\textmd{in}>-1$~$k_BT$. Regarding the surface adsorption, it is called strong if the local minimum located at the hydrogel surface is $V_\textmd{eff}^\textmd{surf}<-1$~$k_BT$, weak if $0>V_\textmd{eff}^\textmd{surf}>-1$~$k_BT$ and metastable when $V_\textmd{eff}^\textmd{surf}>0$.

\begin{figure}[ht!]
  \centering
  \includegraphics[width=\linewidth]{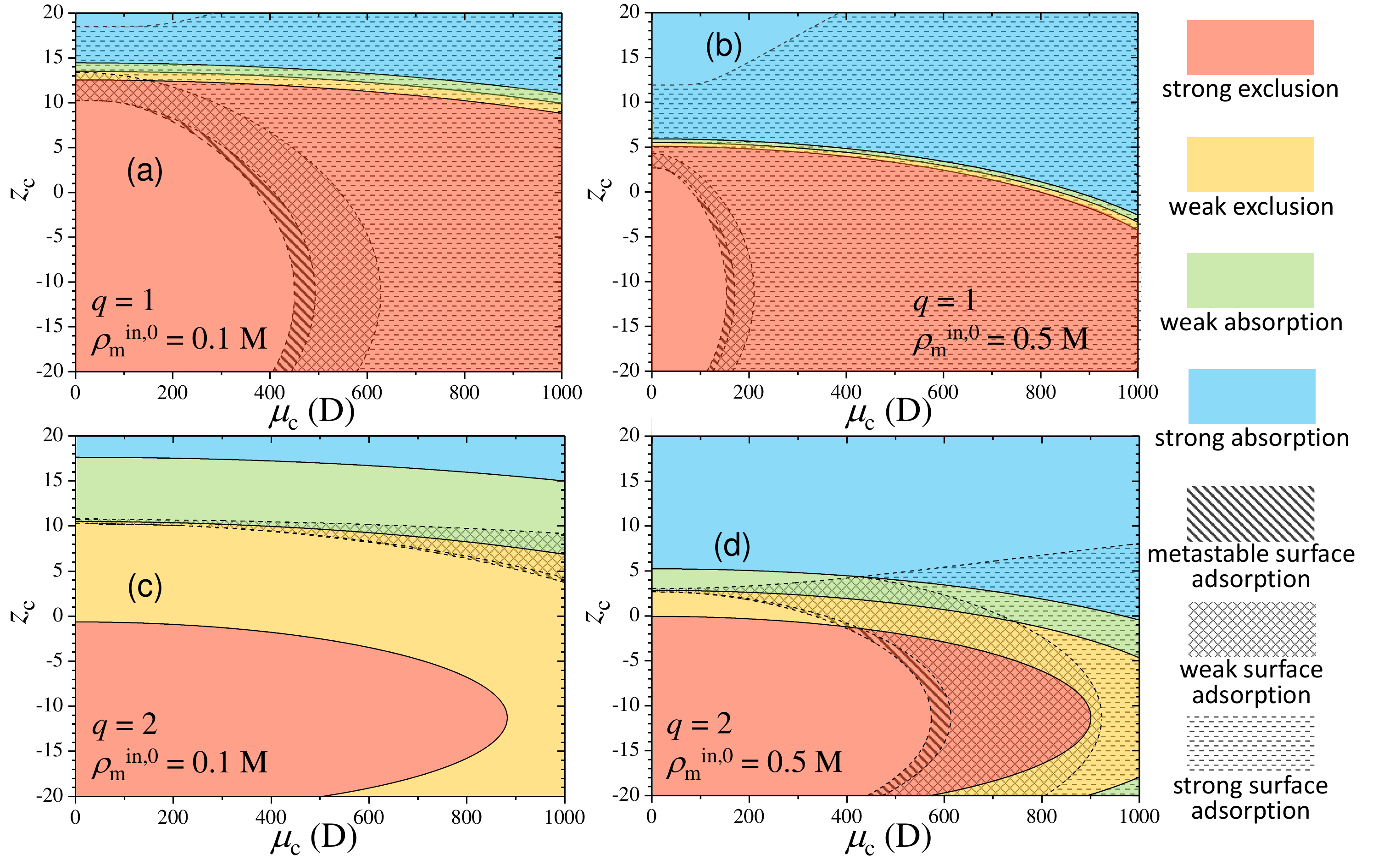}
  \caption{$z_\textmd{c}-\mu_\textmd{c}$ state diagrams for two different hydrogel charge densities, $\rho_\textmd{m}^{in,0}=0.1$~M and $0.5$~M, and for two swelling states, swelling ratio $q=1$ (collapsed) and $q=2$ (swollen). Solid colors represent the strength of exclusion/absorption inside the hydrogel membrane, whereas the different dashed/striped areas denote the regions of the state diagrams where metastable and stable surface adsorption occurs.}
  \label{fig:fig3}
\end{figure}

In order to illustrate the role of the cosolute charge and dipole moment on the absorption/adsorption equilibrium state, we represent in Figure~\ref{fig:fig3} the above mentioned classification into $z_\textmd{c}-\mu_\textmd{c}$ state diagrams for two swelling states (collapsed and swollen) and two hydrogel density charges. Solid colored regions identify the different internal absorption/exclusion strengths, whereas striped and dashed regions refer to the three different surface adsorption states: metastable, weak and strong adsorption.

We first discuss the results obtained for a collapsed hydrogel ($q=1$). For $\rho_\textmd{m}^{\textmd{in},0}=0.1$~M (Figure~\ref{fig:fig3}(a)), strong exclusion dominates over the most part of the diagram due to the very large steric repulsion exerted by the polymer network in such collapsed state (of about $10$~$k_BT$). Only for strong oppositely charged cosolutes weak exclusion and weak absorption states may be found. For $z_\textmd{c}>15$, the electrostatic attraction dominates over steric exclusion, leading to strong absorption states. For large values of $\mu_\textmd{c}$, the dipolar attraction induces strong surface adsorption, whereas metastable and weak surface adsorption are found for small and moderate $\mu_\textmd{c}$. In fact, in the region of strong internal exclusion, increasing $\mu_\textmd{c}$ at fixed $z_\textmd{c}$ leads to metastable, weak and finally strong surface adsorption. However, it should be emphasized that surface adsorption states  appear even for non-polar cosolutes due to the competition between steric exclusion and electrostatic attraction (see the region above $z_\textmd{c}>10$ for $\mu_\textmd{c}=0$). This kind of states have been found, for instance, in the interaction between cross-linked poly(acrylic acid) hydrogels and oppositely charged poly-L-lysine peptides of large molecular weight~\cite{Bysell2006}. Peptides smaller than the typical network pore size are unable to penetrate the hydrogel and become concentrated at its external interface.

It is interesting to note the existence of a reentrant region. For instance, at fixed $\mu_\textmd{c}=600$~D, if we move upward by increasing $z_\textmd{c}$ from $-20$, the state shifts from strong adsorption, to weak adsorption, and finally to strong adsorption again. The explanation of this effect relies on the interplay between the different contributions to the effective potential. For this particular choice of $\mu_\textmd{c}$, the dipolar attraction induces a potential minimum at the hydrogel interfaces of about $-4$~$k_BT$. For $z_\textmd{c}=-1$, the combination of the strong internal exclusion and the dipolar attraction leads to strong surface adsorption, with a minimum at the interface of about $-1.4$~$k_BT$. By decreasing the cosolute charge to $z_\textmd{c} =-12$, the enhancement of the monopolar electrostatic repulsion reduces the potential well at the interface to about $-0.8$~$k_BT$ (weak surface adsorption). If we increase even more the cosolute charge to $z_\textmd{c} =-20$, both the monopolar repulsion and Born attraction increase, but the monopolar term grows proportionally to $z_\textmd{c}$, whereas Born attraction does it as $z_\textmd{c}^2$ (see Eqs.~\ref{Velec} and \ref{VBorn} in the Methods Section). This enhanced Born attraction does not compensate the internal exclusion, but reinforces the surface adsorption, inducing again strong adsorption states (potential well at the interfaces of about $-1.2$~$k_BT$).

By increasing the hydrogel charge density to $\rho_\textmd{m}^{\textmd{in},0}=0.5$~M (see Figure~\ref{fig:fig3}(b)), a similar state diagram is achieved, but now the regions are shifted. The region of strong internal absorption now appears for smaller cosolute charges due to the enhancement of monopolar and Born attractions. States associated to weak internal exclusion and weak internal absorption become confined to a very thin region of the diagram, which means that the system rapidly goes from strong exclusion to strong internal absorption with only a small increase of $z_\textmd{c}$. The region of strong surface adsorption becomes also larger due to the enhancement of the dipolar attraction, and extends from smaller values of $\mu_\textmd{c}$.

Important changes are observed in the diagrams for $q=2$ (swollen state). In this case, the steric repulsion is greatly reduced (steric barrier of $0.96$~$k_BT$), so the diagram becomes more sensitive to the electrostatic interactions. As a consequence of the very weak steric preclusion, the state points of strong exclusion are now confined in the region $z_\textmd{c} < 0$. In other words, only electrostatic repulsion can prevent the cosolute from diffusing inside the hydrogel. As the swollen hydrogel also has a smaller charge density, the effective interactions become in general weaker. This effect is clearly observed in the plot for $\rho_\textmd{m}^{\textmd{in},0}=0.1$~M (Figure~\ref{fig:fig3}(c)), where the regions of weak internal exclusion and weak internal absorption become greatly expanded. Only for oppositely highly charged cosolutes, the strong internal absorption may be reached. Interestingly, the competition between monopolar repulsion and Born attraction leads  to another state reentrance. Due to the weak interaction forces, strong surface adsorption does not occur for this swelling configuration, and metastable and weak surface adsorption becomes reduced to a very small region. 


Increasing the hydrogel density charge to $\rho_\textmd{m}^{\textmd{in},0}=0.5$~M (Figure~\ref{fig:fig3}(d)) emphasizes the electrostatic effects. As a consequence, the region of strong absorption extends to smaller values of $z_\textmd{c}$, and the regions of weak and strong surface adsorption become expanded. Increasing $\mu_\textmd{c}$ leads to stronger dipolar attractions, which gradually pushes the state from metastable to strong adsorption. The re-entrance appearing in the adsorption states also has a similar explanation, although in this case the dipolar term is also involved in the energy balance. 

\begin{figure}[ht!]
  \centering
  \includegraphics[width=\linewidth]{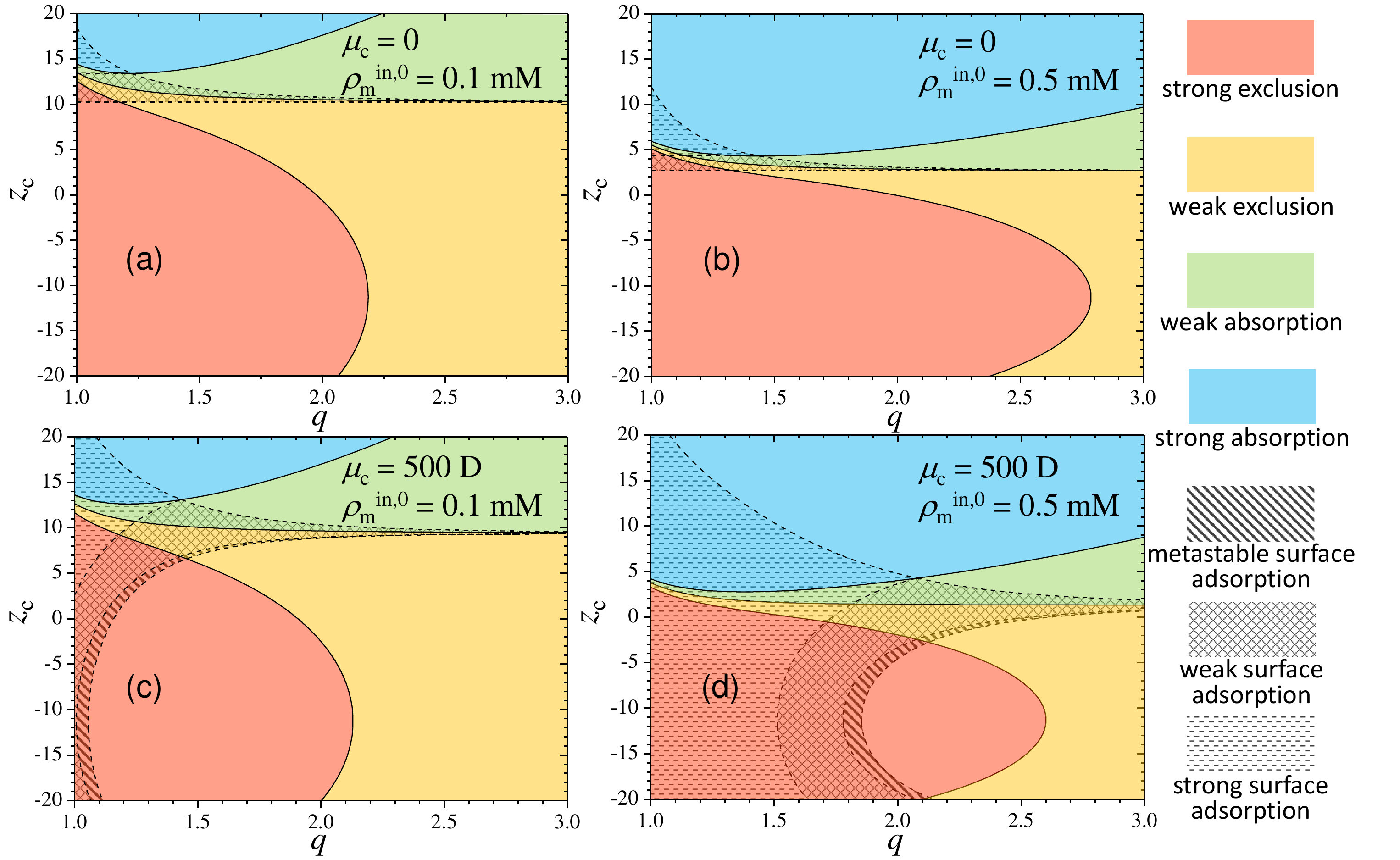}
  \caption{$z_\textmd{c}-q$ state diagrams for two different hydrogel charge densities, $\rho_\textmd{m}^{in,0}=0.1$~M and $0.5$~M, and for two cosolute dipole moments, $\mu_\textmd{c}=0$ (apolar) and $\mu_\textmd{c}=500$~D (polar). Solid colors represent the strength of exclusion/absorption inside the hydrogel membrane, whereas the different dashed/striped areas denote the regions of the state diagrams where metastable and stable surface adsorption occurs.}
  \label{fig:fig4}
\end{figure}

The effect of the hydrogel swelling is studied in Figure~\ref{fig:fig4}, where four $z_\textmd{c}-q$ diagrams obtained for two hydrogel density charges and for non-polar and polar cosolutes are depicted. Figure~\ref{fig:fig4}(a) shows the results for $\rho_\textmd{m}^{\textmd{in},0}=0.1$~M and $\mu_\textmd{c}=0$. As observed, the region of strong internal exclusion dominates for likely-charged cosolutes. If the hydrogel swells, the internal exclusion becomes weaker as the steric repulsive barrier decreases. For shrunken states, the diagram is dominated by strong exclusion, with an small corner of strong absorption that arises for highly oppositely charged cosolutes. For swollen states, only weak exclusion and weak absorption takes place. 

Interestingly, for swollen hydrogels, the transition between weak internal absorption and weak internal exclusion becomes independent on the swelling ratio, $q$. This behavior can be rationalized in terms of the balance between the significant energetic contributions of the effective potential (Eqs.~\ref{Donnan}, \ref{h}, \ref{VBorn} and \ref{Vsteric} in the Methods Section). In this limit, the monopolar term inside the hydrogel network scales with $q$ as $\beta e \psi\sim z_\textmd{c}z_\textmd{m}\rho_\textmd{m}^{\textmd{in},0}/(2\rho_\textmd{s}q^3)$, the Born attraction becomes $\beta V_\textmd{Born}\sim -\kappa_\textmd{bulk}\lambda_B\rho_\textmd{m}^{\textmd{in},0}z_\textmd{c}^2/(8\rho_\textmd{s}(1+\kappa_\textmd{bulk}R_\textmd{c})^2q^3)$, whereas the steric term tends to $\beta V_\textmd{steric}\sim (1+R_\textmd{c}/R_\textmd{m})^2\phi_\textmd{p}^\textmd{in}/q^3$.  Since all contributions to $V_\textmd{eff}$ decrease asymptotically with $q^{-3}$ in the limit of large $q$, the transition between weak absorption and weak exclusion (which occurs when $V_\textmd{eff}=0$) becomes independent of $q$. For the particular values of Figure~\ref{fig:fig4}(a), this transition happens for $z_\textmd{c}=10.2$. 

Strong and weak surface adsorption states also emerge, although they are confined in the region close to the collapsed state. In this limit, a small region
of weak adsorption and strong exclusion is found, which means that cosolutes tend to adsorb onto the hydrogel interfaces. Additionally, strong surface adsorption states lie almost completely inside the strong internal absorption region, which implies a partitioning between surface adsorption and internal absorption. For this particular case, the formation of an adsorbed monolayer of cosolute onto the hydrogel interface can preclude the dynamics of internal absorption processes for concentrated cosolute suspensions. 

Increasing the hydrogel charge density to $\rho_\textmd{m}^{\textmd{in},0}=0.5$~M (see Figure~\ref{fig:fig4}(b)) leads to similar qualitative behavior, with the exception that the region for strong internal absorption is expanded to smaller $z_\textmd{c}$ due to the enhancement of the electrostatic attraction. The regions of strong and weak surface adsorption are also shifted to smaller values $z_\textmd{c}$ for the same reason.

Finally, Figure~\ref{fig:fig4}(c) and Figure~\ref{fig:fig4}(d) illustrate the location of the absorption/adsorption states for a polar cosolute with $\mu_\textmd{c}=500$~D. In both cases, the enhanced Born attraction displaces the internal exclusion to larger negative values of $z_\textmd{c}$. However, the most significant change observed for polar cosolutes is the enlargement of the regions of surface adsorption states. For larger values of $q$, the value of the dipolar attraction found for such swollen states is not high enough to induce surface adsorption.

\subsection{Encapsulation kinetics}

We solved the DDFT differential equations (Eqs.~\ref{ddft2}--\ref{Dc} of Methods section) to determine the time evolution of the cosolute concentration. Within this theoretical framework, the cosolute uptake kinetics may be described in terms of the cosolute diffusion coefficient inside the hydrogel, the cosolute concentration and $V_\textmd{eff}(r)$.

In the initial stage, cosolutes are uniformly distributed outside the hydrogel. For subsequent times, the diffusion process leads to a gradual increase of the concentration inside the internal hole until the finally equilibrium state is achieved, {\it i.e.} when the concentration inside the hydrogel hole matches the initial bulk concentration. For $r=5b$ the distance is far enough from the hydrogel to suppose that the cosolute concentration can be approximated by its bulk value. To solve the time-dependent DDFT differential equations, distances were rescaled by $l_0=1$~nm, and time by $\tau_0=l_0^2/D_0=1.076 \cdot 10^{-9}$~s, where $D_0=k_BT/(6\pi \eta R_c)=2.45\cdot 10^{-10}$~m$^2/$s is the cosolute diffusion coefficient in the bulk solution. In order to shorten the computation time of the numerical resolution, a non-uniform spacial grid was used to sample the distance $r$: a smaller grid size is required at both hydrogel interfaces, where the gradients of the diffusion constant and the effective interaction are higher, whereas larger size intervals are employed in the regions inside and outside the hydrogel. In our calculations we chose $\Delta r_\textmd{min}=0.02l_0$. On the other hand, a time step of $\Delta t=10^{-4}\tau_0$ was used in all the calculations. This value is smaller that $(\Delta r_\textmd{min})^2/(2D_0)$ in order to fulfill the stability condition and prevent the appearance of unphysical saw-tooth waves.

\begin{figure}[ht!]
  \centering
  \includegraphics[width=0.5\linewidth]{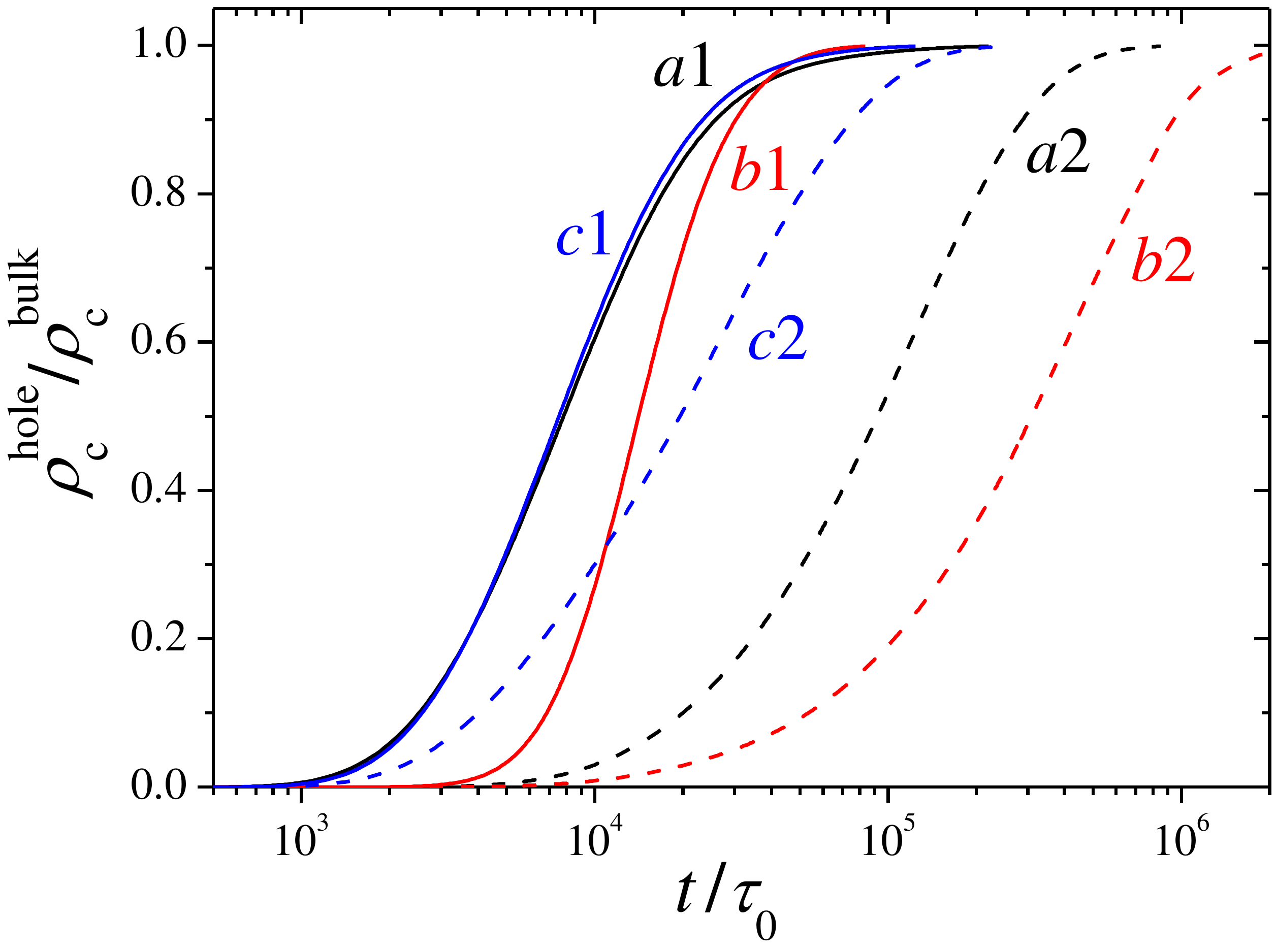}
  \caption{Normalized cosolute concentration inside the hydrogel hole for different conditions. (\textit{a}1)-(\textit{a}2) $\rho_\textmd{m}^\textmd{in}=0.1$~M, $z_\textmd{c}=0$, $\mu_\textmd{c}=0$, $q=3$ and $q=1.5$, respectively.  (\textit{b}1)-(\textit{b}2) $\rho_\textmd{m}^\textmd{in}=0.5$~M, $q=1.5$, $\mu_\textmd{c}=0$, $z_\textmd{c}=8$ and $z_\textmd{c}=-2$, respectively.  (\textit{c}1)-(\textit{c}2) $\rho_\textmd{m}^\textmd{in}=0.5$~M, $q=2$, $z_\textmd{c}=0$, $\mu_\textmd{c}=800$~D and $\mu_\textmd{c}=1300$~D, respectively.}
  \label{fig:fig5}
\end{figure}

During the process, integration of the density profile allows the calculation of the total number of molecules trapped inside the core, in the polymer network, and outside the hydrogel. Figure~\ref{fig:fig5} shows the average cosolute concentration inside the internal hole, defined as
\begin{equation}
\rho_\textmd{c}^{\textmd{hole}}(t)=4\pi\int_0^{a-2\delta}\rho_\textmd{c}(r,t)r^2dr,
\end{equation}
normalized by the bulk concentration for different values of $q$, $z_\textmd{c}$ and $\mu_\textmd{c}$. In all cases, the concentration inside is zero during the initial stage because cosolute particles need some time to diffuse across the hydrogel membrane from the bulk. The duration of this initial lag time depends on the swelling state and the value of the effective interactions. For instance, it is longer for collapsed hydrogels since the repulsive steric barrier is higher and the diffusion coefficient is smaller, and decreases as the electrostatic attraction is enhanced. After this initial delay time, $\rho_\textmd{c}^{\textmd{hole}}(t)$ experiences an accelerated growth that saturates when the cosolute distribution approaches the final equilibrium state.

In the next three sections we explore the effects of $q$, $z_\textmd{c}$ and $\mu_\textmd{c}$ on the encapsulation kinetics. In these three studies, the cosolute radius, bulk cosolute concentration and bulk salt concentration were fixed to $R_\textmd{c}=1$~nm, $\rho_\textmd{c}^\textmd{bulk}=0.001$~M, and $\rho_\textmd{s}=0.1$~M, respectively.

\subsubsection{Effect of the hydrogel swelling}

To explore how the swelling state of the hollow hydrogel affects the dynamics of cosolute encapsulation, a set of calculations was performed for different values of $q$, keeping fixed the rest of parameters. We considered a neutral and apolar cosolute hydrogel ($z_\textmd{c}=0$ and $\mu_\textmd{c}=0$), and a hydrogel of charge density (in the collapsed state) given by $\rho_\textmd{c}^{\textmd{in},0}=0.1$~M.

\begin{figure}[ht!]
  \centering
  \includegraphics[width=0.5\linewidth]{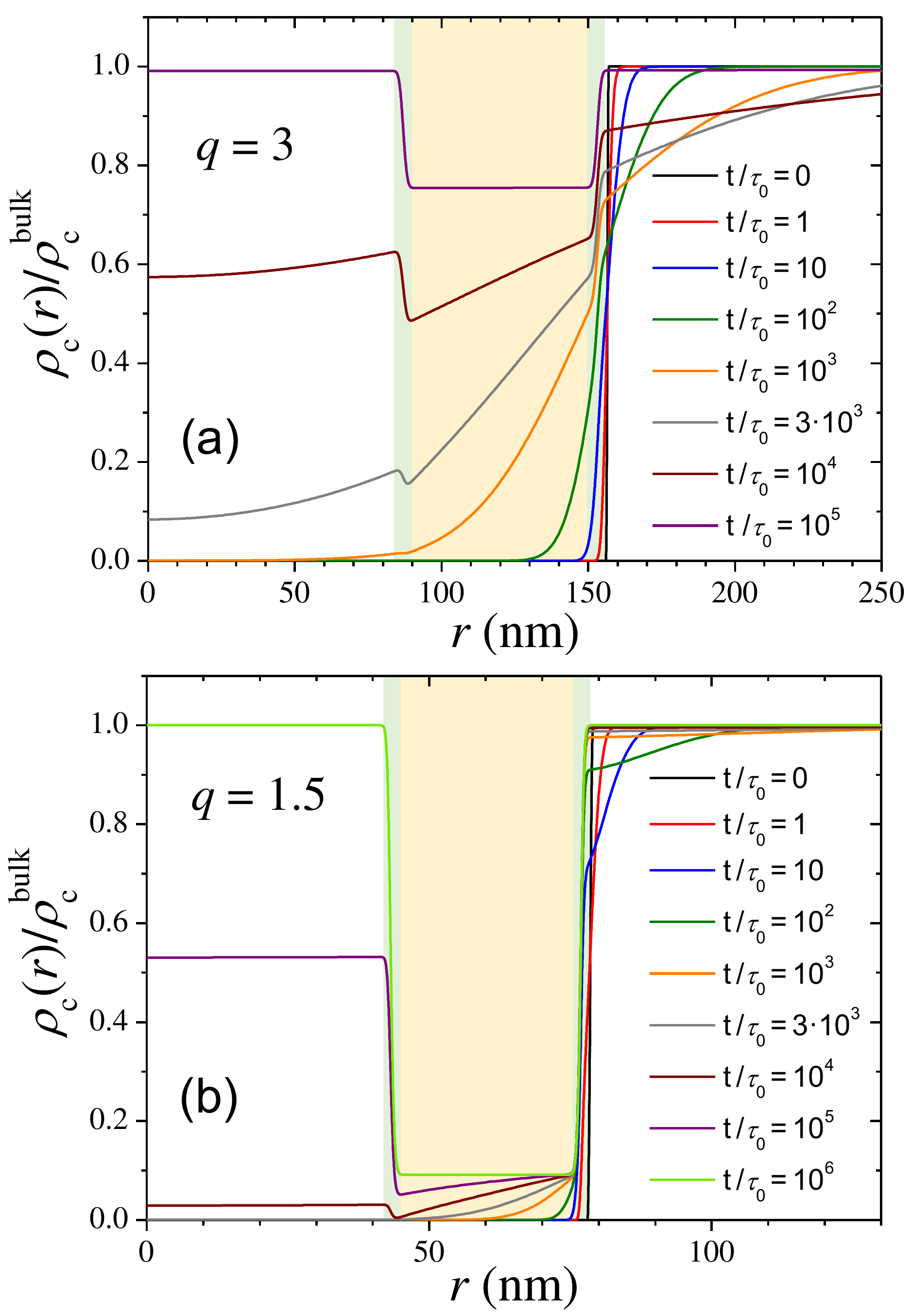}
  \caption{Time evolution of the cosolute density profile for (a) $q=3$ and (b) $q=1.5$. In both cases, $z_\textmd{c}=0$, $\mu_\textmd{c}=0$, and $\rho_\textmd{c}^{\textmd{in},0}=0.1$~M.}
  \label{fig:fig6}
\end{figure}

Figure~\ref{fig:fig6} illustrates the time evolution of the normalized cosolute density profile for a swollen state, $q=3$ (top panel), and for a partially collapsed state, $q=1.5$ (bottom panel). The hydrogel membrane and its two interfaces are depicted in colored background. In the initial stage, cosolute molecules are outside the hydrogel. As the time evolves, they diffuse through the hydrogel membrane until the internal hole is filled at the bulk concentration. For $q=3$ the cosolute concentration is only slightly depleted in the final equilibrium state. However, this depletion effect becomes much more important for $q=1.5$, where the steric barrier is about $2.4$~$k_BT$, leading to very sharp concentration gradients at both interfaces. Also the cosolute diffusion process through the hydrogels takes longer for $q=1.5$. The reduction of the in-diffusion rate when hydrogel collapses can also be clearly appreciated in Figure~\ref{fig:fig5}, where the time evolution of the cosolute concentration inside the hydrogel hole is plotted for both swelling states (curves $a1$ and $a2$, respectively). This is caused by the combination of two effects: (1) cosolute molecules need more time to overcome the steric barrier located at the external interface of the hydrogel, and (2) the diffusion process inside the membrane is slower due to the steric and hydrodynamic obstruction effects (see Eq.~\ref{Dc} in the Methods Section). Therefore, this kinetic regime can be regarded as \textit{steric-precluded diffusion}.

\begin{figure}[ht!]
  \centering
  \includegraphics[width=0.5\linewidth]{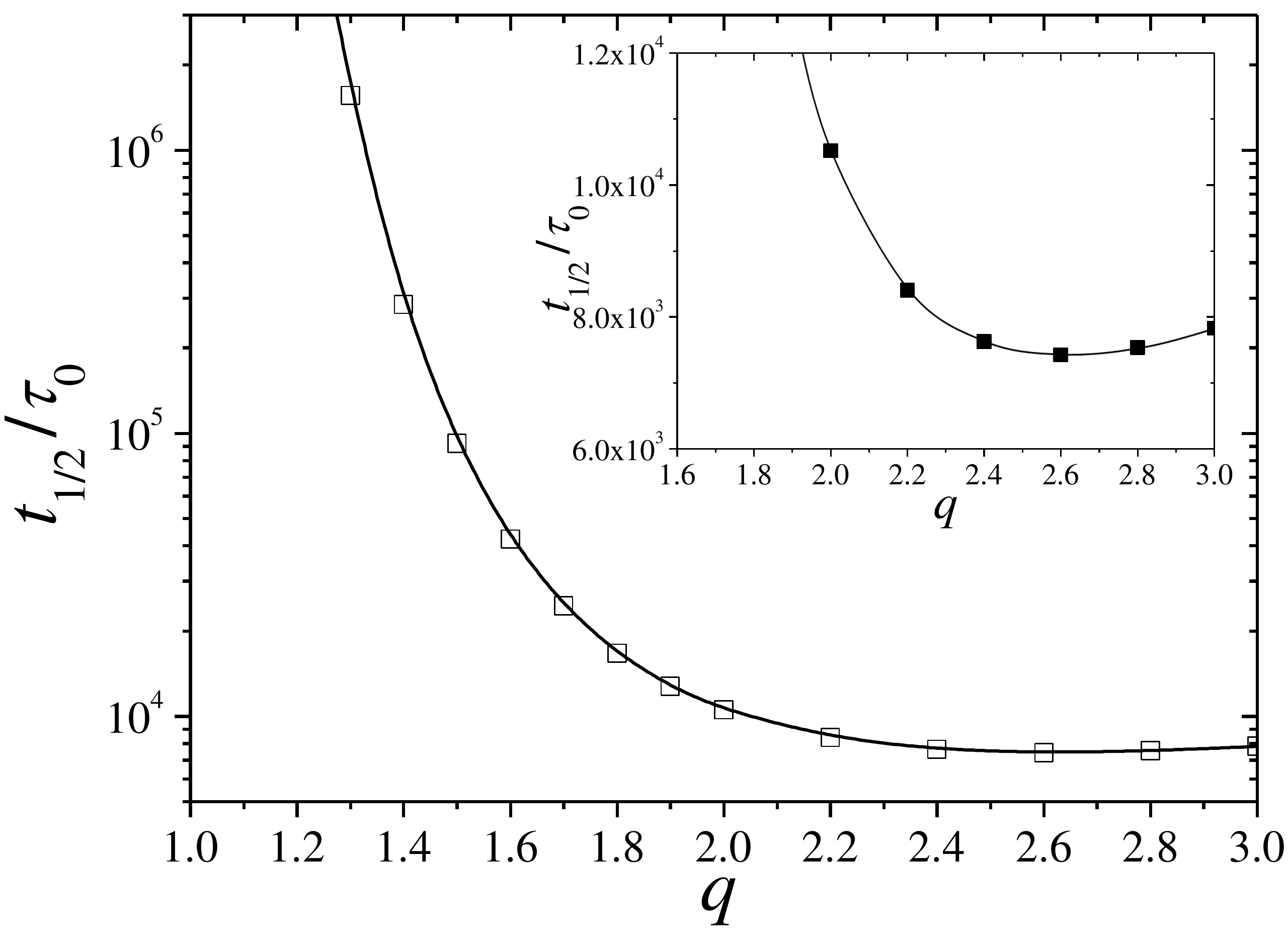}
  \caption{Time required to fill the internal hole at half of the bulk concentration, $t_{1/2}$, as a function of the swelling ratio, $q$. Inset: blow-up for $1.6<q<3.0$. }
  \label{fig:fig7}
\end{figure}

The time required to fill the hydrogel hole with half of the bulk cosolute concentration, $t_{1/2}$, is plotted in Figure~\ref{fig:fig7}. This plot confirms that, in general, $t_{1/2}$ shows a strong increase as the hydrogel collapses, which is a clear signature of the enhancement of the excluded-volume interactions in this steric-precluded regime.

It is also interesting to note the existence of a non-monotonic behavior in the region of highly swollen states, where $t_{1/2}$ decreases from $q=3$ to $q=2.6$ (see inset of Figure~\ref{fig:fig7}). In this regime, the polymer volume fraction is so diluted that obstruction and steric effects are negligible, so cosolutes diffuse across the hydrogel membrane almost like in the bulk. However, for $q=3$ the hydrogel is also more expanded, so the cosolute needs additional time to travel along the whole extension of the network, and so increasing $t_{1/2}$.

\subsubsection{Effect of cosolute charge}

In a second set of calculations, the effect of the cosolute net charge ($z_\textmd{c}$) on the encapsulation kinetics is analyzed for apolar cosolutes ($\mu_\textmd{c}=0$) and for an hydrogel with $q=1.5$ and $\rho_\textmd{m}^{\textmd{in},0}=0.5$~M. The cosolute charge is increased from $z_\textmd{c}=-4$ (strong electrostatic repulsion of cosolute to polymer) to $z_\textmd{c}=+8$ (strong electrostatic attraction of cosolute to polymer).

\begin{figure}[ht!]
  \centering
  \includegraphics[width=0.5\linewidth]{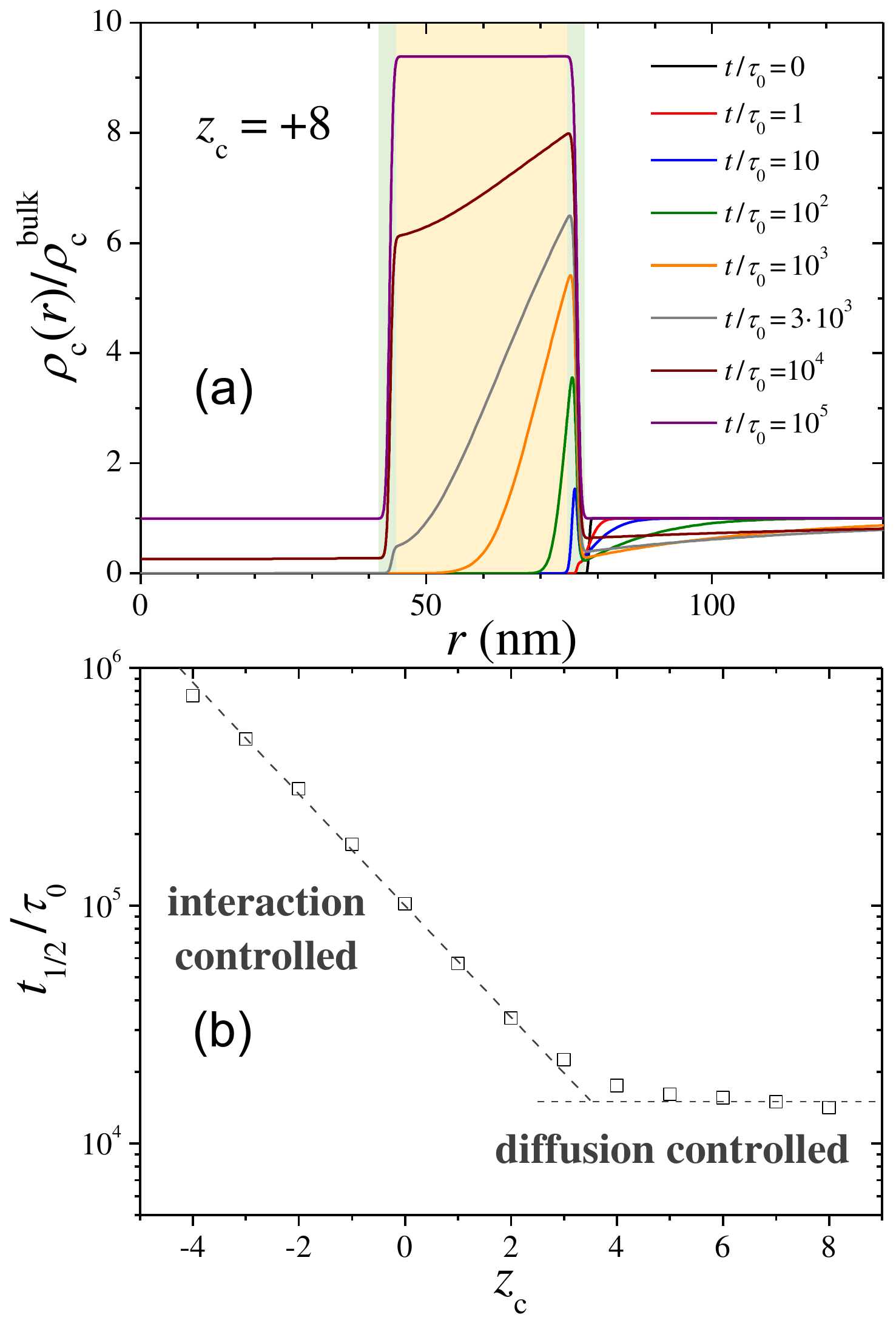}
  \caption{(a) Time evolution of the cosolute density profile for cosolutes oppositely charged to the polymer shell, with $z_\textmd{c}=+8$. (b) Time required to reach half of the bulk concentration inside the hole, $t_{1/2}$, as a function of $z_\textmd{c}$. In both plots $\mu_\textmd{c}=0$, $q=1.5$ and $\rho_\textmd{m}^{\textmd{in},0}=0.5$~M.}
  \label{fig:fig8}
\end{figure}

Figure~\ref{fig:fig8}(a) depicts the time evolution of the cosolute density profile for $z_\textmd{c}=+8$. In this case, the cosolute is strongly electrostatically attracted to the hydrogel polymer network. The concentration inside the hydrogel membrane starts to grow very fast from its external side due to the absorption of cosolutes from the bulk. Progressively, the cosolute concentration in the membrane tends to flatten. At the final equilibrium state, the internal hole is filled with the same concentration than the outside bulk, but there is a significant high concentration absorbed inside the corona of the hollow hydrogel. Here, cargo uptake is driven by the electrostatic absorption at the hydrogel membrane, and this strong electrostatic attraction significantly facilitates the encapsulation process. This kind of electrostatically-driven absorption process has been experimentally observed, for instance, in the upload of proteins such as lysozime and human serum albumin into charged carboxymethylated poly(hydroxyethyl methacrylate) hydrogels~\cite{Garrett1998}. 

Conversely, the encapsulation process for likely charged cosolutes ($z_\textmd{c}<0$) becomes very slow. In this case, the combination of electrostatic and steric repulsion strongly reduces the uptake rate, leading to an important depletion of molecules in the hydrogel membrane. This phenomenon is also illustrated in Figure~\ref{fig:fig5} (curves $b1$ and $b2$), where a significant reduction of the encapsulation kinetics occurs when $z_\textmd{c}$ is varied from $+8$ to $-2$. This decrease of the in-diffusion rate has been found in the uploading of anionic dyed cargos across hollow hydrogel membranes in water purification applications due to the electrostatic rejection induced by previously adsorbed cargos~\cite{Tripathi2014}. 

A more detailed study of the effect of $z_\textmd{c}$ on the encapsulation kinetics can be observed in Figure~\ref{fig:fig8}(b), where the time to reach half the equilibrium loading is plotted against $z_\textmd{c}$. Two well-defined kinetic regimes are clearly appreciated in this plot. On the one hand, for $z_\textmd{c} \geq +4$, the combination of Born solvation and monopolar electrostatic attraction dominate over the rest of repulsive terms, leading to strong cosolute absorption ($V_\textmd{eff}^\textmd{in}<-1$~$k_BT$). In this regime, the cosolute is so attracted to the polymer network that rapidly crosses the external interface of the hydrogel. Here, the kinetic behavior does not change with the cosolute charge, as it is fully controlled by the diffusion time across the hydrogel membrane: \textit{diffusion-controlled regime}. On the other hand, for $z_\textmd{c} < +4$, the steric repulsion start to affect the dynamics. This steric effect is strengthen even more by the monopolar electrostatic repulsion for $z_\textmd{c} < 0$, leading to a larger repulsion as $z_\textmd{c}$ becomes more negative. In order to diffuse inside the hole, the cosolute molecules must be able to surpass this effective repulsive barrier. This leads to $t_{1/2}$ that grows exponentially with the cosolute charge, in the so-called \textit{interaction-controlled regime}.

\begin{figure}[ht!]
	\centering
	\includegraphics[width=0.5\linewidth]{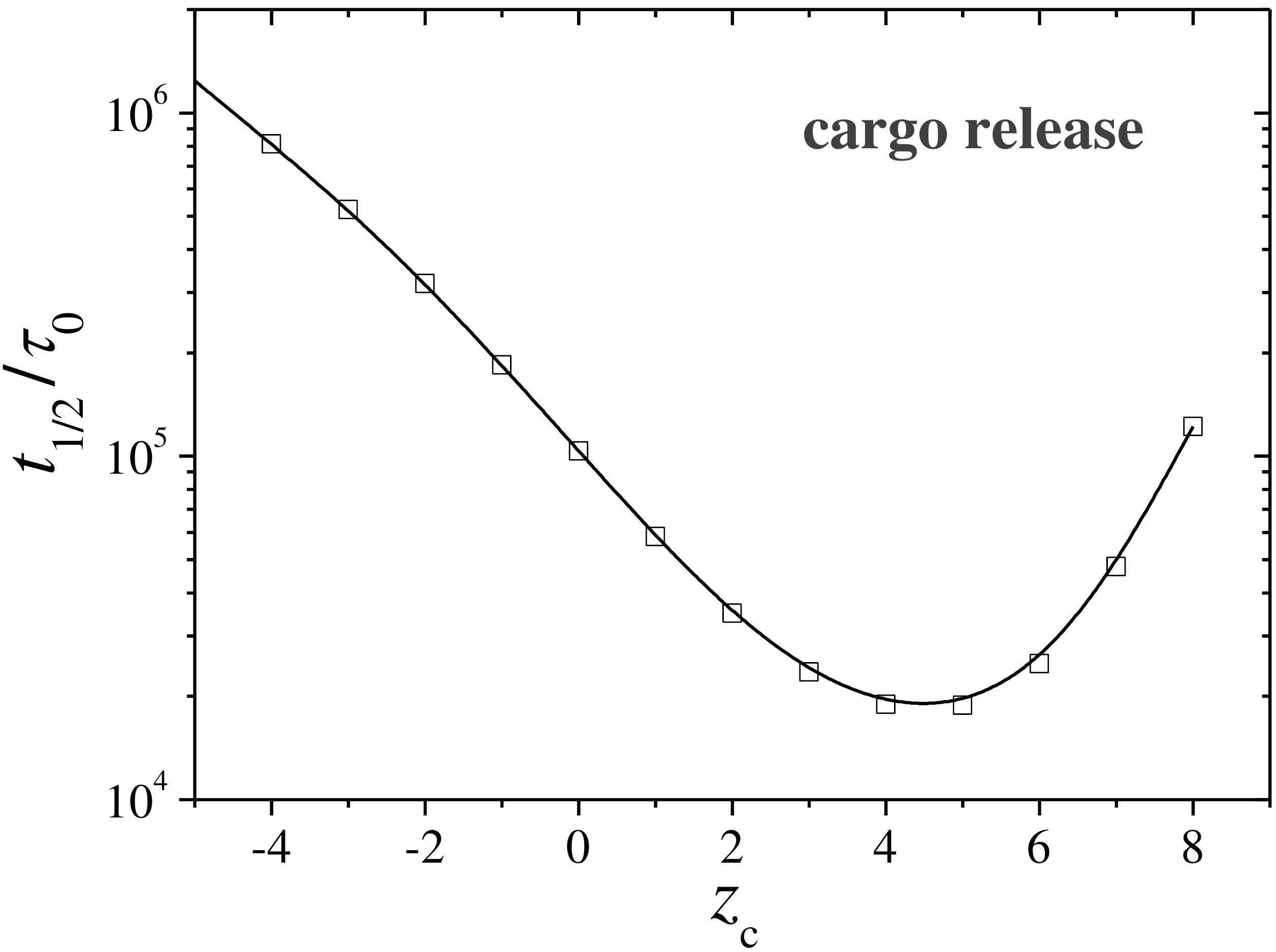}
	\caption{Time required to release half of the encapsulated cargo, $t_{1/2}$, as a function of $z_\textmd{c}$, for $\mu_\textmd{c}=0$, $q=1.5$ and $\rho_\textmd{m}^{\textmd{in},0}=0.5$~M.}
	\label{fig:fig9}
\end{figure}

However, it should be emphasized that the release kinetics from the internal hole to the bulk suspension is not just a simple reverted process of the uptake. In fact, it shows a very different qualitative outcome. This is clearly illustrated in Figure~\ref{fig:fig9}, which depicts the time to release half of the encapsulated cosolute ($t_{1/2}$) as a function of the cosolute charge. In particular, when the cosolute and the hydrogel are oppositely charged, the release of the cargo contained in the internal hole becomes seriously precluded due to the electrostatic trapping within the polymer shell during escape. For like charged solutes the polymer constitutes a large barrier for escape. These effects lead both to a decrease of the diffusion rate with higher absolute charge that can strongly inhibit the cargo release, and to the appearance of a non-monotonic behavior of the release half-time as a function of the cosolute charge that is not observed in the encapsulation process. This reduction on the release rate caused by the electrostatic attraction has been experimentally observed in the kinetics of positively charged doxorubicin molecules encapsulated into anionic superparamagnetic hollow hybrid nanogels at high pH~\cite{Chiang2013}.


\subsubsection{Effect of cosolute electric dipole moment}

In the last set of calculations, we explore the effect of the cosolute electric dipole moment on the encapsulation kinetics for the particular case of neutral cosolutes ($z_\textmd{c}=0$), and for a swollen hydrogel with $q=2$ and $\rho_\textmd{m}^{\textmd{in},0}=0.5$~M. According to the state diagram shown in Figure~\ref{fig:fig3}(d), increasing the polarity (here dipole) of the cosolute particle for this choice of parameters leads to different absorption and adsorption states: from weak exclusion to strong absorption, and from metastable adsorption to strong stable adsorption.

\begin{figure}[ht!]
  \centering
  \includegraphics[width=\linewidth]{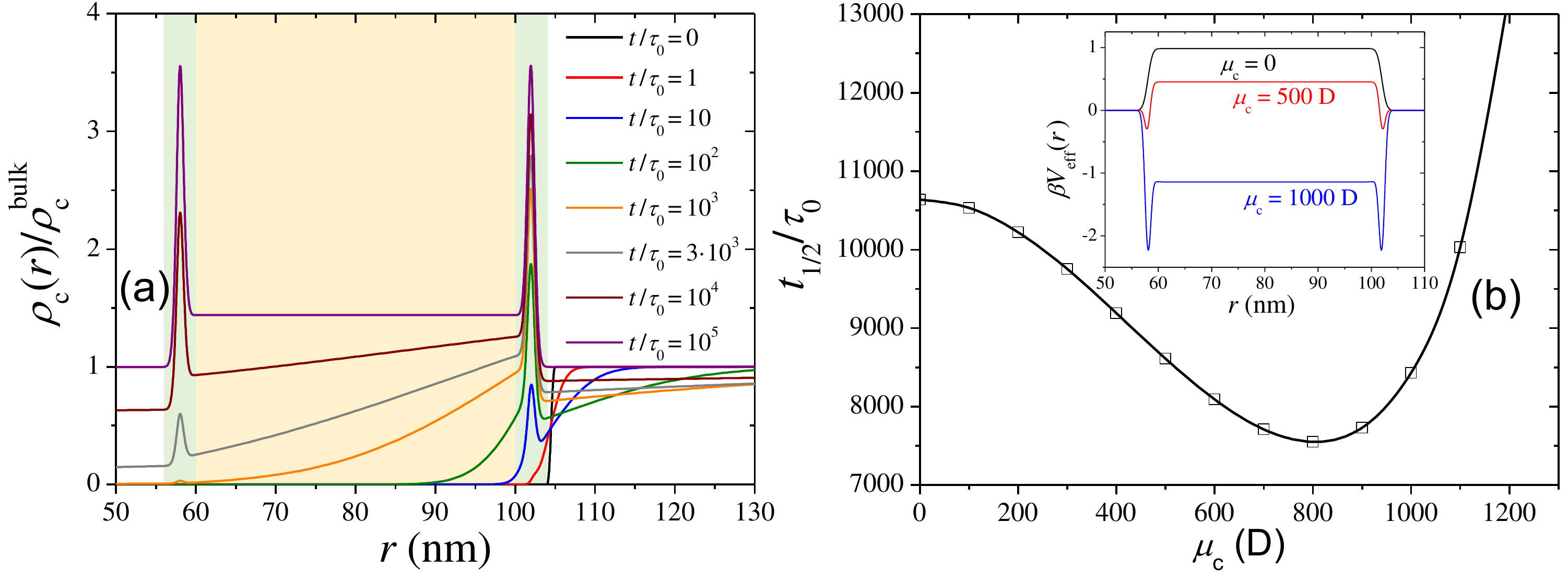}
  \caption{(a) Time evolution of the cosolute density profile for polar cosolutes with $\mu_\textmd{c}=800$~D. (b) Time required to reach half of the bulk concentration inside the hole, $t_{1/2}$, as a function of $\mu_\textmd{c}$. In both plots $z_\textmd{c}=0$, $q=2$ and $\rho_\textmd{m}^{\textmd{in},0}=0.5$~M.}
  \label{fig:fig10}
\end{figure}

Figure~\ref{fig:fig10}(a) plots $\rho_\textmd{c}(r,t)$ for $\mu_\textmd{c}=800$~D. As observed, a peak in the cosolute concentration shows up at the external interface of the hydrogel for very short times. This surface adsorption process is driven by the dipolar attractive term (Eq.~\ref{Vdip} in the Methods Section), which generates an attractive potential well at both hydrogel interfaces of about $-1.32$~$k_BT$. As the encapsulation process proceeds and the cosolute propagates across the membrane, a second concentration peak located at the internal interface develops at longer diffusion times for the same reason. In the final equilibrium state, the dipolar contribution to the Born attraction yields weak cosolute absorption inside the hydrogel network.

Again, a clearer picture of the effect of the dipole moment can be gained in Figure~\ref{fig:fig10}(b), where $t_{1/2}$ is plotted against $\mu_\textmd{c}$. Interestingly, $t_{1/2}$ shows a non-monotonic dependence as the result of the interplay between different energetic terms of the effective interaction. For small values of $\mu_\textmd{c}$, an increase of $\mu_\textmd{c}$ enhances the Born solvation attraction, which accelerates the in-diffusion process. However, for $\mu_\textmd{c} > 800$~D, the dipolar attraction becomes very important, giving rise to two surface-adsorption minima that deepen by further increasing $\mu_\textmd{c}$ (see inset in Figure~\ref{fig:fig10}(b)). Both minima prevent the diffusion process, especially at the external layer where the cosolute becomes strongly adsorbed and accumulates in time, impeding the diffusion of more molecules through the membrane. As a result of this, $t_{1/2}$ shows a quite significant growth with $\mu_\textmd{c}$. This result can also be appreciated in Figure~\ref{fig:fig5} (curves $c1$ and $c2$), where $\rho_\textmd{c}^\textmd{hole}(t)$ shows a faster time evolution for $\mu_\textmd{c}=800$~D than for $1200$~D. We call this effect \textit{adsorption-hindered diffusion}. 

\section{Conclusions}
\label{sec:conclusions}

In summary, we studied the equilibrium states and the non-equilibrium encapsulation kinetics of charged cosolutes in the presence of a hollow charged hydrogel in salty suspensions for a wide range of solute charges and electric dipole moments, and for hydrogels with different charge density, covering swelling ratios ranging from $q=1$ (collapsed state) to $q=3$ (swollen state).

Our results indicate that the swelling state of the hollow hydrogel plays a very important role on determining the equilibrium cosolute distribution and the uptake dynamics, due to the excluded-volume repulsion induced by the polymer network and the osmotic pressure exerted by the excess counterions inside the hydrogel. The strong cosolute obstruction achieved close to the collapsed state emphasize volume-exclusion inside the hydrogel membrane, promotes the appearance of surface adsorption states and strongly hinders the encapsulation kinetics (\textit{steric-precluded} regime). These steric effects are reinforced by the electrostatic repulsion for likely charged cosolutesm leading to an \textit{interaction-controlled} kinetic regime, in which the encapsulation kinetics slows down exponentially with the cosolute charge.

Conversely, steric and osmotic exclusion becomes counterbalanced by the electrostatic attraction for the case of cosolutes oppositely charged than the polymer. In fact, cosolutes with high values of $z_\textmd{c}$ tend to be strongly absorbed into the polymer network of the hydrogel. In this regime, the encapsulation process is significantly accelerated, and becomes almost completely independent on $z_\textmd{c}$. In other words, the cosolute uploading is controlled by the diffusion time required to cross the hydrogel membrane (\textit{diffusion-controlled} regime). However, it should be emphasized that in this regime, the uptake mainly occurs in hydrogel shell, and not in the internal hole.

The non-uniform charge distribution of the cosolute particle has also important implications. In particular, surface adsorption onto both hydrogel interfaces are promoted when increasing the cosolute electric dipole moment due to the enhancement of the effective cosolute-hydrogel dipolar attraction. We report a non-monotonic behavior of the encapsulation kinetics in terms of $\mu_\textit{c}$: For small values $\mu_\textit{c}$, the increase of Born solvation attraction with $\mu_\textit{c}$ emphasizes its absorption in the polymer network of charged hydrogels, increasing the encapsulation rate. However, for high enough $\mu_\textit{c}$, the kinetics enters the so-called \textit{adsorption-hindered} regime, in which surface adsorption at the external layer impedes the in-diffusion, which gives rise to a slow down of the encapsulation kinetics.

The model presented here can be easily extended to investigate the inverse case of cargo {\it release} kinetics which could exhibit very different kinetic behavior, {\it i.e.}, would not be a simple time reversal of the uptake process. In particular, when the cosolute and the hydrogel are oppositely charged, the release of the cargo contained in the internal hole becomes seriously precluded due to the electrostatic `trapping' within the polymer shell during escape. For like charged solutes, on the other hand, the polymer constitutes a large barrier for escape from the void. These effects leads both to a decrease of the diffusion rate with higher absolute charge and hence to the appearance of a non-monotonic behavior of the release time as a function of the cosolute charge that is not observed in the encapsulation process. A detailed discussion will follow in a future study. 

Moreover, in many of the applications of hydrogels particles as nanoreactors or drug delivery vectors, the kinetics of the solute uptake (or release) may have an important repercussion on the swelling state of the hydrogel~\cite{Moncho-Jorda2016,Kim2017}. Therefore, the generalization of the free-energy functional to incorporate the swelling dynamics in response of the cosolute absorption/exclusion/adsorption is also a fruitful future work. There have been, for instance, peculiar phenomena related to mechanical instabilities near the consolute points~\cite{Gilcreest, Halperin2018,Gorelov1997}. Finally, recent simulations and theoretical predictions performed with hydrophobic cosolutes show that the interplay between hydrophobic adhesion and steric exclusion leads to a maximum in the uploaded cosolute for certain intermediate swelling state~\cite{Perez-Mas2018}. Therefore, including an additional specific polymer-cosolute free-energy contribution in our present model will incorporate the effect of hydrogel bonds and short-range hydrophobic/hydrophilic forces~\cite{Moncho-Jorda2014}.

\section{Methods}
\label{sec:theory}

\subsection{Calculation of the effective hydrogel-cosolute interaction}
\label{sec:effectivepotential}

One of the most important parameters determining the cosolute equilibrium density profiles around a hydrogel particle and its absorption kinetics is the effective hydrogel-cosolute interaction, $V_\textmd{eff}(r)$. In this work, we follow a similar phenomenological representation for this effective pair potential previously used in earlier studies, and assume that it can be split into three additive contributions: electrostatic, osmotic and steric (see Eq.~\ref{Veff}).~\cite{Yigit2017,Adroher-Benitez2017}

$V_\textmd{elec}$ represents the effective electrostatic interaction between the cosolute and the charge distribution of the hydrogel and the ions surrounding it. Neglecting high order multipolar contributions, $V_\textmd{elec}(r)$ can be expressed as the sum of a monopolar term, an attractive orientation-averaged dipolar term, and a Born solvation self-energy 
\begin{equation}
\label{Velec}
V_\textmd{elec}(r)=z_\textmd{c} e \psi(r) +V_\textmd{dip}(r) + \Delta V_\textmd{Born}(r).
\end{equation}
Here, $\psi(r)$ is the electrostatic potential induced by the hydrogel together with its ionic double layer, and $E(r)=-d\psi(r)/dr$ is the corresponding local electric field. Considering a Boltzmann distribution of ions and assuming electroneutrality (realized by the high enough concentration of monovalent salt, as the Debye screening length is much smaller than the polymer shell thickness), $\psi(r)$ takes the form of a Donnan potential~\cite{Yigit2012}
\begin{equation}
\label{Donnan}
\beta e \psi(r)=\ln \left[h(r)+\sqrt{1+h(r)^2}\right]
\end{equation}
where $\beta =1/(k_BT)$ ($T$ is the absolute temperature, and $k_B=1.38\times 10^{-23}$~J/K is the Boltzmann constant), and $h(r)$ is defined as the ratio between the local charge density induced by cosolutes and hydrogel charged monomers, and the bulk concentration of ions ($2\rho_\textmd{s}$) 
\begin{equation}
\label{h}
h(r)=\frac{z_\textmd{m}\rho_\textmd{m}(r)+z_\textmd{c}\rho_\textmd{c}(r)}{2\rho_\textmd{s}}.
\end{equation}
Depending of the sign of the cosolute particle, $z_\textmd{c} e \psi(r)$ can be attractive or repulsive, and always reaches its maximum absolute value in the center of the hydrogel membrane. 

The second term of Eq.~\ref{Velec} represents the dipole interaction between the cosolute permanent dipole moment ($\mu_\textmd{c}$) and the electrostatic mean field generated by the charged hydrogel together with the ionic double layer, $E(r)=-d\psi(r)/dr$. For strong $E(r)$, the dipole tends to align in the same direction than $E(r)$, whereas for low electrostatic fields, it becomes randomly oriented due to thermal fluctuations. The final alignment is a competition between both effects, and can be expressed as a statistical average of the Boltzmann factor, performed over all possible angular orientations. This leads to~\cite{Hill1986,Yigit2017}
\begin{equation}
\label{Vdip}
\beta V_\textmd{dip}(r)=-\ln \left[  \frac{\sinh (\beta \mu_\textmd{c} | E(r) |)}{\beta \mu_\textmd{c} |E(r)| } \right].
\end{equation}
The dipolar term induces an effective attraction in the regions where $E(r)$ reaches its maximum value, namely the internal and external interfaces of the hydrogel membrane. Therefore, cosolutes with high values of $\mu_\textmd{c}$ will tend to get adsorb onto both layers.

The third contribution of the electrostatic term is usually referred as the Born interaction. It reflects the change in the self-energy cost of charging the cosolute inside the charged hydrogel {\it versus} bulk solvent, $\Delta V_\textmd{Born}(r)=V_\textmd{Born}\left(\kappa(r)\right)-V_\textmd{Born}({\kappa_\textmd{bulk}})$. In the Debye-H\"uckel approximation, the leading order of this interaction up to the dipole level is~\cite{Yigit2012,Yigit2017}
\begin{eqnarray}
\label{VBorn}
\beta V_\textmd{Born}(\kappa)&=&\lambda_B\frac{z_\textmd{c}^2}{2R_\textmd{c}\left(1+\kappa R_\textmd{c}\right)}   \\
&+& \frac{3\lambda_B\mu_\textmd{c}^2\left(1+\kappa R_\textmd{c}\right)\left(2+2\kappa R_\textmd{c}+\kappa^2R_\textmd{c}^2\right)}{2e^2R_\textmd{c}^3\left(3+3\kappa R_\textmd{c}+\kappa^2R_\textmd{c}^2\right)^2},     \nonumber
\end{eqnarray}
where $\lambda_B=e^2/(4\pi\eps_r\eps_0k_BT)$ is the Bjerrum length, $\kappa (r)=(4\pi \lambda_B ( \rho_+(r) + \rho_-(r) + z_\textmd{m}^2\rho_\textmd{m}(r))^{1/2}$ is the local inverse Debye screening length at position $r$, and $\kappa_\textmd{bulk}=(8\pi \lambda_B\rho_\textmd{s})^{1/2}$ \cite{Angioletti-Uberti2018}. In the definition of $\kappa(r)$, $\rho_+(r)$ and $\rho_-(r) $ are the number density of positive and negative salt ions, respectively, and we have assumed the specific case of monovalent salts. Due to the higher value of the screening constant inside the hydrogel network, the Born free energy represents an attractive interaction that pulls the charged cosolute inside the hydrogel network. It should be noted that deviations from Eq.~\ref{VBorn} can arise for strongly charged cosolutes due to non-linear effects beyond the Debye-H\"uckel approximation, or for highly asymmetric distribution of charges within the cosolute due to the appearance of quadrupolar and higher order contributions to the electrostatic energy.

In addition to the electrostatic interaction, the cosolute also experiences a volume work against the osmotic pressure exerted by the ions inside the hydrogel network~\cite{Adroher-Benitez2017}. Within the ideal gas approximation, the effective osmotic repulsion inside the hydrogel reads as
\begin{equation}
\beta V_\textmd{osm}(r)= \frac{4}{3}\pi R_\textmd{c}^3 \left[ \rho_+(r) + \rho_-(r) -2 \rho_\textmd{s} \right].
\end{equation}

It is important to remark that, for strongly charged polymer networks, certain amount of counterions may be condensed onto the polymer chains, so the real concentration of mobile counterions should be corrected by the Manning theory or other improved models that treat the counterion condensation phenomenon~\cite{Manning1969,Zeldovich1999,Dobrynin2005}. As the hydrogel charge density considered in this work is small enough to keep the Manning parameter below unity, we neglect this effect.

Lastly, but not less important, the third term of the right hand side of Eq.~\ref{Veff} represents the excluded-volume (or steric) repulsion caused by the polymer chains onto the cosolute particle. It is formally given by $\beta V_\textmd{steric}=-\ln (v_\textmd{avail}/v)$, where $v$ is the total volume of the hydrogel and $v_\textmd{avail}$ is the actual available volume (not excluded by the polymers) for a cosolute particle of radius $R_\textmd{c}$. $v_\textmd{avail}$ can be calculated for certain morphologies of the hydrogel network. In particular, if the internal polymer network is modelled by an assembly of randomly oriented straight polymer chains of radius $R_\textmd{m}$, the steric repulsion is given by the following analytical expression by~\cite{Ogston1958,Bosma2000,Moncho-Jorda2013}
\begin{equation}
\label{Vsteric}
\beta V_\textmd{steric}(r)=-\left( 1+ \frac{R_\textmd{c}}{R_\textmd{m}}\right)^2\ln (1-\phi_\textmd{p}(r)),
\end{equation}
where $\phi_\textmd{p}(r)$ is the local polymer volume fraction. This term is strongly dependent on the swelling state of the hydrogel. It is negligible for swollen hydrogels but leads to very high repulsive steric barriers near the collapsed state, therefore hindering the cosolute in-diffusion across the hydrogel. This model for the steric repulsion should be regarded as a first approximation that fulfills the right limiting behavior, namely $V_\textmd{steric}\rightarrow 0$ for $\phi_\textmd{p}\rightarrow 0$, and $V_\textmd{steric}\rightarrow \infty$ for $\phi_\textmd{p}\rightarrow 1$. In general, this interaction is determined by the geometrical constrains induced by the polymer fibers, and so may depend on the particular morphology of the cross-linked polymer network, the polymer flexibility, and on the existence of polymer fluctuations, which can reduce the obstruction exerted by the polymers, especially in the limit of collapsed networks~\cite{Kanduc2018}. We also note that we neglected in this work the action of a mean-field attractive contribution,~\cite{Adroher-Benitez2017} for example stemming from a mildly hydrophobic or dispersion interaction between solutes and polymer. While in general relevant, we avoided this additional parameter for now as it would in principle only scale the steric interaction (11) up and down and not add more qualitative insight to the already convoluted kinetic findings. 

The total effective interaction is connected to the equilibrium cosolute distribution through the relation $\rho_\textmd{c}^\textmd{eq}(r)=\rho_c^\textmd{bulk}e^{-\beta V_\textmd{eff}(r)}$. Note that the electrostatic potential itself depends on the equilibrium density ({\it via} the Donna term) and thus the equation must be solved self-consistently. Therefore, by examining the form of $V_\textmd{eff}(r)$, the degree of absorption inside the core and at the external corona of the hollow hydrogel can be determined, and the location where the cosolute will preferentially partition predicted. However, as it will be shown in the following section, not only the equilibrium properties are affected by $V_\textmd{eff}(r)$, but also the dynamic properties arising in non-equilibrium conditions.

\subsection{Dynamic Density Functional Theory for the cosolute encapsulation kinetics}
\label{sec:DDFT}

To investigate the time evolution of the cosolute concentration, we make use of Dynamic Density Functional Theory (DDFT) as our theoretical framework~\cite{Marconi1999,Wu2007}. This method describes how an initial non-equilibrium density profile of cosolute particles evolves in time to finally reach the equilibrium state, in the presence of an external potential. In contrast to the ideal diffusion equation, DDFT also considers the effect of the cosolute-cosolute and cosolute-hydrogel interactions and the position dependence of the diffusion coefficient that occurs when the cosolute diffuses through the polymer network.

Since the cosolute cannot be created or destroyed in the system (mass conservation), its diffusion follows the  continuity equation: 
\begin{equation}
\frac{\partial \rho_\textmd{c}(\vec{r},t)}{\partial t}=-\nabla \cdot J_\textmd{c}(\vec{r},t),
\end{equation}
where $J_\textmd{c}(\vec{r},t)$ denotes the space and time-dependent net flux. According to DDFT, this flux is proportional to the gradient of the cosolute chemical potencial ($\tilde{\mu}_\textmd{c}$),
\begin{equation}
J_\textmd{c}(\vec{r},t)=-D_\textmd{c}(\vec{r})\rho_\textmd{c}(\vec{r},t)\nabla (\beta \tilde{\mu}_\textmd{c}(\vec{r},t)).
\end{equation}

Due to the spherical symmetry of the hydrogel, both equations can be written in terms of the distance to the hydrogel center, $r$:
\begin{equation}
\label{ddft2}
\frac{\partial \rho_\textmd{c}}{\partial t}=-\frac{1}{r^2}\frac{\partial}{\partial r}\Big( r^2   J_\textmd{c} \Big) \ \ \ , \ \ \
J_\textmd{c}(r,t)=-D_\textmd{c}(r)\rho_\textmd{c}(r,t)\frac{\partial (\beta \tilde{\mu}_\textmd{c})}{\partial r}. 
\end{equation}

The main assumption of this theory relies on the approximation that the above defined non-equilibrium space and time dependent chemical potential can be deduced from the functional derivative of the equilibrium free energy~\cite{Evans1992,Hansen2013}
\begin{equation}
\tilde{\mu}_\textmd{c}(\vec{r},t)=\frac{\delta F[\rho_\textmd{c}(\vec{r},t)]}{\delta \rho_\textmd{c}(\vec{r},t)}.
\end{equation}

In equilibrium, $\tilde{\mu}_\textmd{c}$ becomes time and position independent, the net flux becomes zero, and so the cosolute concentration becomes time independent, as expected.

In order to have a complete theory, we need to find a fair approximation for the equilibrium free energy functional. In this work, we propose the following expression~\cite{Angioletti-Uberti2014}:
\begin{eqnarray}
\label{free_energy}
\beta F[\rho_\textmd{c}(\vec{r})]&=&\int \rho_\textmd{c}(\vec{r}) \left[\ln ( \rho_\textmd{c}(\vec{r})\Lambda_\textmd{c}^3)-1\right] d\vec{r}  \nonumber \\
&+&\int \rho_\textmd{c}(\vec{r})\beta V_\textmd{eff}(r)d\vec{r}+\int\beta f_\textmd{HS}(r) d\vec{r},
\end{eqnarray}
where the integrals are performed over the entire volume of the system, and $\Lambda_\textmd{c}$ is the thermal wave length of the cosolute. The first term of Eq.~\ref{free_energy} corresponds to the ideal contribution. The second one accounts for the interaction of the cosolutes with the effective mean-field potential  $V_\textmd{eff}(r)$ induced by the hydrogel and the ionic cloud (which includes the electrostatic, osmotic and steric contributions deduced in the previous section). The third term accounts for the short-range repulsion among the cosolute molecules. We neglect any specific attraction between cosolute molecules. We thus approximate the cosolutes by hard spheres with certain effective radius $R_\textmd{c}$, and approximate the free-energy density by the Carnahan-Starling expression~\cite{Hansen2013}
\begin{equation}
\beta f_{HS}(r)=\frac{4\phi_\textmd{c}(r)-3\phi_\textmd{c}(r)^2}{(1-\phi_\textmd{c}(r))^2}\rho_\textmd{c}(r),
\end{equation}
where $\phi_\textmd{c}(r)=\frac{4}{3}\pi R_\textmd{c}^3\rho_\textmd{c}(r)$ is the local cosolute volume fraction. It should be noted that, for large cosolute concentrations and large dipoles, dipole-dipole interactions between the solutes should also be explicitly considered in the theory. In this work we approximate that dipole interactions only act on the one-body level with the mean-field electric external field, $E(r)$. Note that the next order correction would be a rescaling of the local dielectric constant~\cite{andelman} which is an interesting issue for future work.

The theory presented above requires that the solvent and ion degrees of freedom relax very fast compared to the cosolute diffusion. Since ionic species and water molecules are in general very small compared to the size of cosolutes such as proteins and other biomolecules, it is a good approximation to assume that ions and solvent molecules distribute instantaneously around the cosolutes during the absorption process.

Finally, before solving the differential equations, it is of crucial importance to estimate the cosolute diffusion coefficient inside the polymer network of the hydrogel. Indeed, inside hydrogels or other porous media where the pore diameters, inter-fiber spacings, or other dimensions of the microstructure are comparable to the size of a diffusing macromolecule, the diffusion coefficient becomes smaller than that in bulk solution, and the percentage of reduction increases with molecular size. The steric restrictions on the positions that can be occupied by a finite-sized cosolute caused by the presence of polymers leads to an increased hydrodynamic drag.

Although many different approximate models can be found in the literature regarding this issue~\cite{Amsden1998, masaro1999physical}, here we make use of an expression for the diffusion coefficient that takes into account obstruction and hydrodynamic effects. It reads as~\cite{Johnson1996,Bosma2000}
\begin{equation}
\label{Dc}
\frac{D_\textmd{c}}{D_0}=\frac{e^{-0.84\alpha^{1.09}}}{1+\sqrt{2\alpha}+2\alpha/3}
\end{equation}
where $D_0$ is the cosolute diffusity in the bulk solution. The dependence on the polymer volume fraction within the hydrogel comes through parameter $\alpha$, which can be approximated by $\alpha=\beta V_\textmd{steric}=-(1+R_\textmd{c}/R_\textmd{m})^2\ln(1-\phi_\textmd{p})$. The numerator on Eq.~\ref{Dc} accounts for the steric obstruction caused by the polymers, which partially inhibits the free diffusion of the cosolute, as many displacements are not allowed. On the other hand, the denominator is the Brinkman's equation for the hydrodynamic retardation effect~\cite{Brinkman1949}. According to this model, the polymer chains around the solute restrict the solvent in its motion, which leads to an enhanced friction between the cosolute and the solvent. The final cosolute diffusivity is the combination of both effects, and leads to a strong reduction of $D_\textmd{c}$ as the hydrogel network approaches the collapsed state. It should be emphasized that the mobility of hydrophobic or polar cosolutes can become relatively complex in denser (especially collapsed) networks where it may be significantly reduced due to specific adsorption onto the polymer chains, which leads to a non-viscous, activated-diffusion process.~\cite{Kanduc2018} A similar `trapping' effect may arise for strongly charged polymers when electrostatic cosolute condensation may occur.~\cite{lifson} For these situations, Eq.~\ref{Dc} needs to be generalized to correctly account for such phenomena.

The position dependent diffusion coefficient, $D_\textmd{c}(r)$, is obtained considering the dependence of the polymer volume fraction with $r$, $\phi_\textmd{p}(r)$, given by Eq.~\ref{phipr}.

In order to solve the DDFT differential equations, we need to specify three boundary conditions. The first condition establishes that the net flux in the center of the hydrogel must be zero due to the spherical symmetry of the system
\begin{equation}
J_\textmd{c}(r=0,t)=0 \ \ \ \ \ \ \ \forall t
\end{equation}
For a very diluted suspension of hydrogel particles, the cosolute concentration far away from the hydrogel must be given by the corresponding bulk value. This is the second boundary condition:
\begin{equation}
\rho_\textmd{c}(r\rightarrow\infty,t)=\rho_\textmd{c}^\textmd{bulk}
\end{equation}
The third condition specifies the initial distribution of cosolute. As we are interested on studying the cosolute loading, we can assume that at time $t=0$ all the cosolute molecules are uniformly distributed outside the hydrogel
\begin{equation}
\label{t0}
\rho_\textmd{c}(r)=\left\{\begin{array}{l@{\quad}l}
0 & r<b \\
\rho_\textmd{c}^\textmd{bulk} & r>b
\end{array}  \right. 
\end{equation}

\begin{acknowledgement}

AMJ thanks the Spanish ``Ministerio de Econom\'{\i}a y Competitividad (MINECO), Plan Nacional de Investigaci\'{o}n, Desarrollo e Innovaci\'{o}n Tecnol\'{o}gica (I + D + i)'' (Project FIS2016-80087-C2-1-P), the European Regional Development Fund (ERDF) and the program ``Visiting Scholars'' funded by the University of Granada. Part of this project (JD) has received funding from the European Research Council (ERC) under the European Union's Horizon 2020 research and innovation programme (grant agreement Nr. 646659). Finally, computer resources provided by PROTEUS (Instituto Carlos I de F\'{\i}sica Te\'{o}rica y Computacional, University of Granada) are gratefully acknowledged.

\end{acknowledgement}

%
%
%

\providecommand{\latin}[1]{#1}
\providecommand*\mcitethebibliography{\thebibliography}
\csname @ifundefined\endcsname{endmcitethebibliography}
  {\let\endmcitethebibliography\endthebibliography}{}

\end{document}